\documentclass[fleqn,usenatbib]{mnras}

\usepackage{amsmath}
\usepackage[varg]{txfonts}

\usepackage[T1]{fontenc}

\DeclareRobustCommand{\VAN}[3]{#2}
\let\VANthebibliography\thebibliography
\def\thebibliography{\DeclareRobustCommand{\VAN}[3]{##3}\VANthebibliography}

\usepackage{graphicx}	% Including figure files
\usepackage{natbib}
\usepackage{enumitem}
\usepackage[flushleft]{threeparttable}
\usepackage{colortbl}
\usepackage[normalem]{ulem}
\usepackage{color}
\usepackage{bm} % bold math
\usepackage{gensymb}
\definecolor{darkblue}{rgb}{0.0,0.0,0.8}
\definecolor{darkred}{rgb}{0.75,0.,0.25}
\definecolor{darkorange}{rgb}{1,0.3,0.0}
\definecolor{darkgreen}{rgb}{0.0,0.6,0.0}
\definecolor{darkpurple}{rgb}{0.8,0.,0.9}
\definecolor{brown}{rgb}{0.65,.16,0.16}
\definecolor{grey}{rgb}{0.4,0.5,0.6}
\definecolor{white}{rgb}{1,1,1}
\definecolor{trolleygrey}{rgb}{0.5, 0.5, 0.5}
\definecolor{lavender}{rgb}{0.835,0.812,0.969}
\definecolor{pastelorange}{rgb}{0.99,0.92,0.82}
\definecolor{pastelblue}{rgb}{0.85,0.93,0.99}

\newcommand{\scf}{{\sc scalefree}}

\newcommand{\gaia}{{\sl Gaia}}
\newcommand{\hst}{{\sl HST}}

\newcommand{\msun}{\rm M_\odot}

\newcommand{\probP}{\text{I\kern-0.15em P}}

\title[Signatures of dark matter subhaloes]{Signatures of dark subhaloes in dwarf spheroidal galaxies: \\ II. Transient and localised kinematic features}

\author[Vitral et al.]{Eduardo Vitral$^{1}$\thanks{Email: eduardo.vitral@roe.ac.uk}\thanks{Royal Society Newton International Fellow},
Jorge Pe\~narrubia$^{1,2,3}$,
Roeland P. van der Marel$^{4,5}$ and Matthew G. Walker$^{6}$
\\
$^1$Institute for Astronomy, University of Edinburgh, Royal Observatory, Blackford Hill, Edinburgh EH9 3HJ, UK\\
$^2$Institute of Corpuscular Physics (IFIC), CSIC–Universitat de València, 46980 Paterna, Valencia, Spain\\
$^3$VALER, Calle Mayor, 83, 1, 12001 Castellón de la Plana, Spain\\
$^4$Space Telescope Science Institute, 3700 San Martin Drive, Baltimore, MD 21218, USA\\
$^5$Center for Astrophysical Sciences, The William H. Miller III Department of Physics \& Astronomy, Johns Hopkins University, Baltimore, MD 21218, USA\\
$^6$McWilliams centre for Cosmology and Astrophysics, Department of Physics, Carnegie Mellon University, Pittsburgh, PA 15213, USA
}

\date{Accepted XXX. Received YYY; in original form ZZZ}

\pubyear{\the\year{}}

\begin{document}
\label{firstpage}
\pagerange{\pageref{firstpage}--\pageref{lastpage}}
\maketitle

% Abstract of the paper
\begin{abstract}
We use controlled $N$-body simulations to investigate whether dark matter subhaloes leave detectable signatures in the internal kinematics of dwarf spheroidal galaxies. 
At each snapshot, we fit a velocity distribution function to stellar line-of-sight velocities and proper motions, construct normalised residual fields for the projected velocity components, and analyse their power-spectra.
We show that subhalo encounters generate a plethora of kinematic features. Amongst the most prominent are rotation-like signals, which arise in our models even though the stellar component is initialised without intrinsic rotation. In line-of-sight velocity alone, several snapshots resemble the classical blueshift--redshift pattern associated with stationary rotation. The plane-of-sky components, however, bring complexity, with alternating regions of expansion and contraction, as well as clockwise and anticlockwise streaming. 
Although we observe coherent streaming globally, these structures depart from typical equilibrium configurations, while their residual power-spectra quantify the characteristic spatial scales involved.
This signal strengthens in models containing more massive subhaloes and becomes difficult to recover for small stellar samples: robust power-spectrum recovery of the full velocity-space signature usually requires $\mathcal{O}\left(10^{4}\right)$ tracers, although a rotation signal alone can be recovered with considerably smaller samples.
These results suggest that dwarf spheroidal velocity fields can retain signatures of subhalo heating. Future spectroscopy, improved \gaia\ astrometry, and chemo-dynamical information should turn these diagnostics into population-level probes of dark matter substructure.
\end{abstract}
% Select between one and six entries from the list of approved keywords.
% Don't make up new ones.
\begin{keywords}
cosmology: dark matter -- Galaxy: structure -- galaxies: kinematics and dynamics -- galaxies: evolution -- galaxies: dwarf -- methods: numerical.
\end{keywords}

%%%%%%%%%%%%%%%%%%%%%%%%%%%%%%%%%%%%%%%%%%%%%%%%%%

%%%%%%%%%%%%%%%%% BODY OF PAPER %%%%%%%%%%%%%%%%%%

\section{Introduction} \label{sec:intro}

The nature of dark matter remains one of the central open questions in modern astrophysics \citep{Balazs+26}. In the standard cosmological picture, galaxies form within extended dark matter haloes that contain a hierarchy of lower-mass, self-bound substructures \citep[see][for a review on this topic, with references therein]{Zavala&Frenk19}. These dark subhaloes are a robust outcome of hierarchical structure formation and encode information about the small-scale behaviour of dark matter. Their abundance, internal structure, survival rate, and spatial distribution therefore provide a direct route to testing the particle nature of dark matter \citep{Colin+02,Bullock&Boylan-Kolchin17,Yang+23,Chiang+25}. The main difficulty, however, is observational: many of these subhaloes are not massive enough to form stars, and are therefore invisible to conventional electromagnetic surveys \citep[e.g.][]{Benitez-Llambay&Frenk20}. Their presence must instead be inferred indirectly, through the gravitational perturbations they imprint on visible tracers \citep{Mao&Schneider98,Zechlin+12,Erkal+16,Bovy+17, Drlica-Wagner+19,Nadler+21,Ballard+24,Delos25,Enzi+25,Cao+25,Tajalli+25}.

Dwarf spheroidal galaxies provide a particularly promising environment in which to search for such perturbations. Their large dynamical mass-to-light ratios \citep{Pryor&Kormendy&Kormendy90,Pace+24}, old stellar populations, and relatively simple baryonic content \citep{Savino+25} make them among the cleanest nearby laboratories for studying dark matter on galactic scales \citep{Simon19}. At the same time, the low stellar masses of these systems imply that even weak gravitational fluctuations sourced by dark substructure can, in principle, leave measurable signatures in the phase-space distribution of their stars \citep{Penarrubia+24}. This motivated the controlled numerical experiments of \citet[][hereafter P25]{Penarrubia+25}, who followed the response of massless stellar tracers embedded in dwarf galaxy dark matter haloes populated by orbiting dark subhaloes. Their simulations showed that repeated subhalo passages drive a gradual expansion of the stellar distribution, while also altering the velocity-dispersion structure of the tracer population.

In \citet*[][hereafter Paper~I]{Vitral+26}, we asked whether these subhalo-driven perturbations could be detected using projected stellar positions alone. The key result was that, although the stellar distribution remains well described at leading order by a smooth density profile, the residual density field contains weak but coherent fluctuations. By comparing the observed stellar counts to those expected from the best-fitting axisymmetric Plummer model, Paper~I constructed a local fluctuation field and analysed it in Fourier space. This approach revealed excess power at characteristic spatial frequencies in simulations with subhaloes, with spectral features that could be described by a constant noise floor plus Voigt-like components. The strength and scale of these features depended on both the parent halo--subhalo mass model and the number of stellar tracers, suggesting that density corrugations in classical dwarf spheroidal galaxies can encode information about the underlying subhalo population.

Projected positions, however, encode only part of the available phase-space information. If dark subhaloes perturb the stellar component through time-dependent gravitational forces, their imprint may also be present in the internal kinematics of the dwarf galaxy (cf. P25). The present paper therefore develops the velocity-space counterpart of Paper~I: we analyse the same class of simulations introduced by P25 and adopted in Paper~I, but now focus on the three projected components of the stellar velocity field. Our goal is not to fit arbitrary velocity maps, but rather to define a physically motivated smooth baseline against which statistically significant departures can be identified, quantified, and ultimately searched for in observational data.

This work is therefore designed to address two main questions: (i) do subhalo interactions leave velocity-space signatures that remain detectable after subtracting a smooth, physically motivated rotating model? And (ii) how do these signatures depend on the parent halo mass, the associated subhalo population, and the number of available stellar tracers? 
These questions are motivated not only by the simulations themselves, but also by observations of dwarf spheroidal galaxies: in Ursa Minor, for instance, \citet{Pace+14} reported two localised secondary kinematic populations whose physical origin could not be uniquely established from line-of-sight data alone, echoing earlier evidence for spatial substructure in this system \citep{Irwin&Hatzidimitriou95,Kleyna+98}. Similar claims have also been made for Sextans, where kinematically distinct components have been reported at different spatial locations \citep{Walker+06,Battaglia+11,Cicuendez&Battaglia18}.
Such cases highlight the need for controlled experiments that identify which mechanisms can generate localised velocity features in otherwise pressure-supported dwarf galaxies. 
They are also directly relevant for interpreting the next generation of dwarf-galaxy kinematic data, for which line-of-sight velocities and plane-of-sky proper motions may provide complementary observables of the same underlying perturbations.

The remainder of this paper is organised as follows. Section~\ref{sec:data} summarises the numerical simulations used in this work, emphasising their connection to P25 and Paper~I. Section~\ref{sec:methods} presents the main ingredients of the velocity-space methodology, while the detailed analytical derivations of the projected velocity moments and Fourier spectra are deferred to the appendices. Section~\ref{sec:results} applies the method to the simulated dwarf galaxies, focusing on the time evolution of the photometric--kinematic misalignment, the rotation-amplitude parameter, the velocity-residual fields, and their dependence on halo mass and tracer number. Section~\ref{sec:discussion} discusses the interpretation of these diagnostics, their relation to equilibrium, and their possible application to observed dwarf galaxies. Finally, Section~\ref{sec:conclusion} summarises our conclusions.

\begin{figure}
\centering
\includegraphics[width=\hsize]{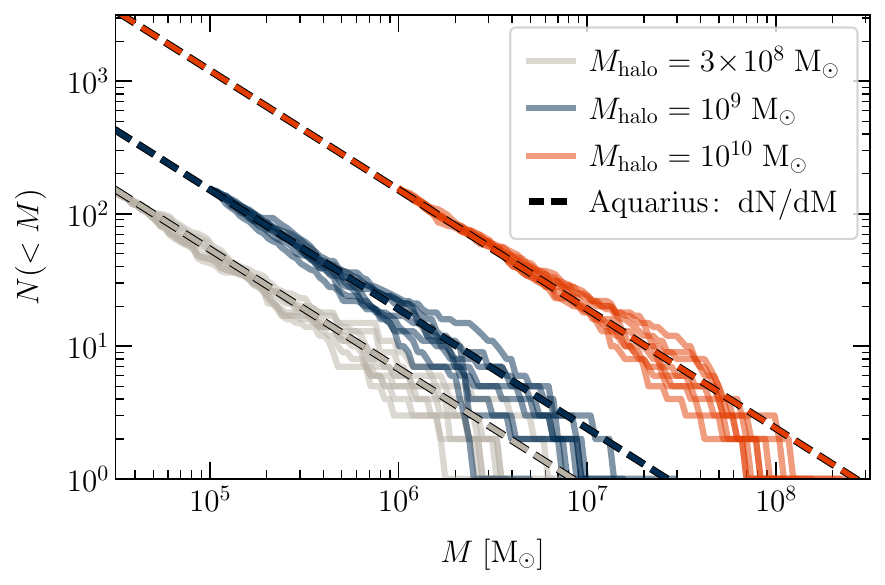}
\caption{\textit{Subhalo masses:} Cumulative number of subhalo masses for the three DM halo models considered in this study, with $M_{\rm halo}~[\msun] = \{ 3~\times~10^{8}, 10^{9}, 10^{10} \}$. 
The subhalo sampling procedure and its dependence on the host halo properties are described in section~2.3 of \protect\cite{Penarrubia+25}.}
\label{fig:subhalo-masses}
\end{figure}

\section{Numerical data}
\label{sec:data}

We analyse the same suite of controlled $N$-body experiments introduced by P25 and adopted in Paper~I. These simulations were designed to isolate the effect of dark subhaloes on the stellar component of dwarf-spheroidal-like galaxies, while deliberately avoiding additional sources of disequilibrium such as tides from a massive host, mergers with luminous companions, or baryonic feedback. This controlled set-up is useful for the present work because it allows any recovered kinematic disturbance to be interpreted relative to a known subhalo population.

The dwarf galaxies are embedded in fixed, spherical dark matter haloes described by \citet{Hernquist90} profiles. We consider the three parent halo masses explored in P25 and Paper~I, namely $M_{\rm halo}=3\times10^{8}~{\rm M}_{\odot}$, $10^{9}~{\rm M}_{\odot}$, and $10^{10}~{\rm M}_{\odot}$. Throughout the paper, we adopt the $10^{9}~{\rm M}_{\odot}$ model as our fiducial example, and explicitly indicate whenever a different halo mass is analysed.
The corresponding halo scale radii were calibrated in P25 to reproduce the characteristic mean densities of Local Group dwarf spheroidals \citep*{Strigari+07_vmax,Penarrubia+08,Kravtsov10,Errani+18}. Because the host potentials are kept fixed, the simulations do not follow the self-consistent response of the smooth dark matter halo to either the stellar component or the subhalo population. This approximation is nevertheless appropriate for the systems considered here, given the large dynamical mass-to-light ratios of dwarf spheroidals \citep{Simon19} and the low individual masses of cosmological subhaloes relative to their host halo \citep[e.g.][]{Springel+08}.

For each parent halo, the subhalo population follows the same cosmologically motivated prescription as in P25. The subhalo mass function is scaled by the host mass and follows the power-law form measured in the Aquarius simulations \citep{Springel+08}, namely ${\rm d}N/{\rm d}M\propto M^{-\alpha}$ with $\alpha=1.9$, over the relative mass interval $M_{\rm sub}/M_{\rm halo}\in[10^{-4},0.03]$,\footnote{P25 showed that the stellar response is driven mainly by the upper end of the subhalo mass spectrum, i.e. by objects with high $M_{\rm sub}/M_{\rm halo}$ ratios. Extending the mass function to smaller ratios would therefore add a large number of low-mass subhaloes, substantially increasing the computational cost, while contributing little to the stellar heating signal or to the physical interpretation of the simulations.} yielding the sampled subhalo masses shown in Figure~\ref{fig:subhalo-masses}.
The spatial number density of subhaloes follows the Hernquist profile of the host halo, yielding an average of approximately $150$ subhaloes per realisation within this mass range. Individual subhaloes are modelled with exponentially truncated \citet*{Navarro+97}-like density profiles \citep[][see also equation~10 of P25]{Errani&Navarro21}, with structural parameters tied to the mean density of the host halo at their pericentre \citep{Errani&Navarro21,Aguirre-Santaella+23}. Their orbits are integrated as non-interacting test particles in the fixed host potential, with initial velocities drawn from an \citet{Osipkov79,Merritt85_df} distribution function chosen to reproduce the expected transition from nearly isotropic inner motions to radially biased outer orbits \citep{Orkney+23}.

The stellar component is initialised as a spherical Plummer tracer population \citep{Plummer1911}, with isotropic initial velocities and an initial scale radius chosen to match the scale of the Sculptor dwarf galaxy \citep{Pace+24}.\footnote{The effect of varying the initial stellar scale radius at fixed halo mass is discussed in detail by P25. In particular, smaller stellar components, associated with denser dark matter haloes, tend to expand more rapidly through subhalo-induced heating.}
The stellar particles are massless tracers of the gravitational potential, which is appropriate for the dark matter-dominated regime targeted by these experiments. In the fiducial realisations, the stellar component contains $N_{\star}=10^{5}$ particles. We also analyse down-sampled catalogues with lower stellar counts, in particular $N_{\star}= \{10^{3}, 10^{4}\}$, in order to assess how the recoverability of the velocity-space signal depends on tracer sampling. Whenever additional stellar counts are used, they are explicitly stated in the corresponding figures and discussion.

Although the stellar distribution is initialised as spherical, subhalo interactions gradually heat and expand the tracer population. P25 showed that this expansion is approximately self-similar, while Paper~I further demonstrated that the projected stellar distribution remains well described, at leading order, by an axisymmetric Plummer profile at each snapshot. We therefore use the axisymmetric Plummer fits from Paper~I as part of the numerical data products adopted here. These fits provide, for every snapshot, the projected centre, scale radius, ellipticity, and position angle of the stellar distribution.

\section{Methods}
\label{sec:methods}

The aim of our methodology is to isolate velocity-space departures from a smooth, physically-motivated description of the stellar component. The procedure has three main steps. First, we fit a projected rotating velocity model to each snapshot, using the axisymmetric Plummer density fits from Paper~I as the spatial backbone of the calculation. Second, following the approach adopted for the stellar density field in Paper~I, we convert the stellar velocities into normalised residual fields, allowing departures to be compared across velocity components and snapshots. Third, we analyse these residual fields in Fourier space to identify the spatial scales on which the kinematic perturbations are most prominent. The detailed projection formulae, likelihood implementation, and spectral fitting choices are given in Appendix~\ref{sec:rotation_models} and Appendix~\ref{sec:spectral_features}. Here, we summarise the main ingredients needed to interpret the results.

\begin{figure*}
\centering
\includegraphics[width=\hsize]{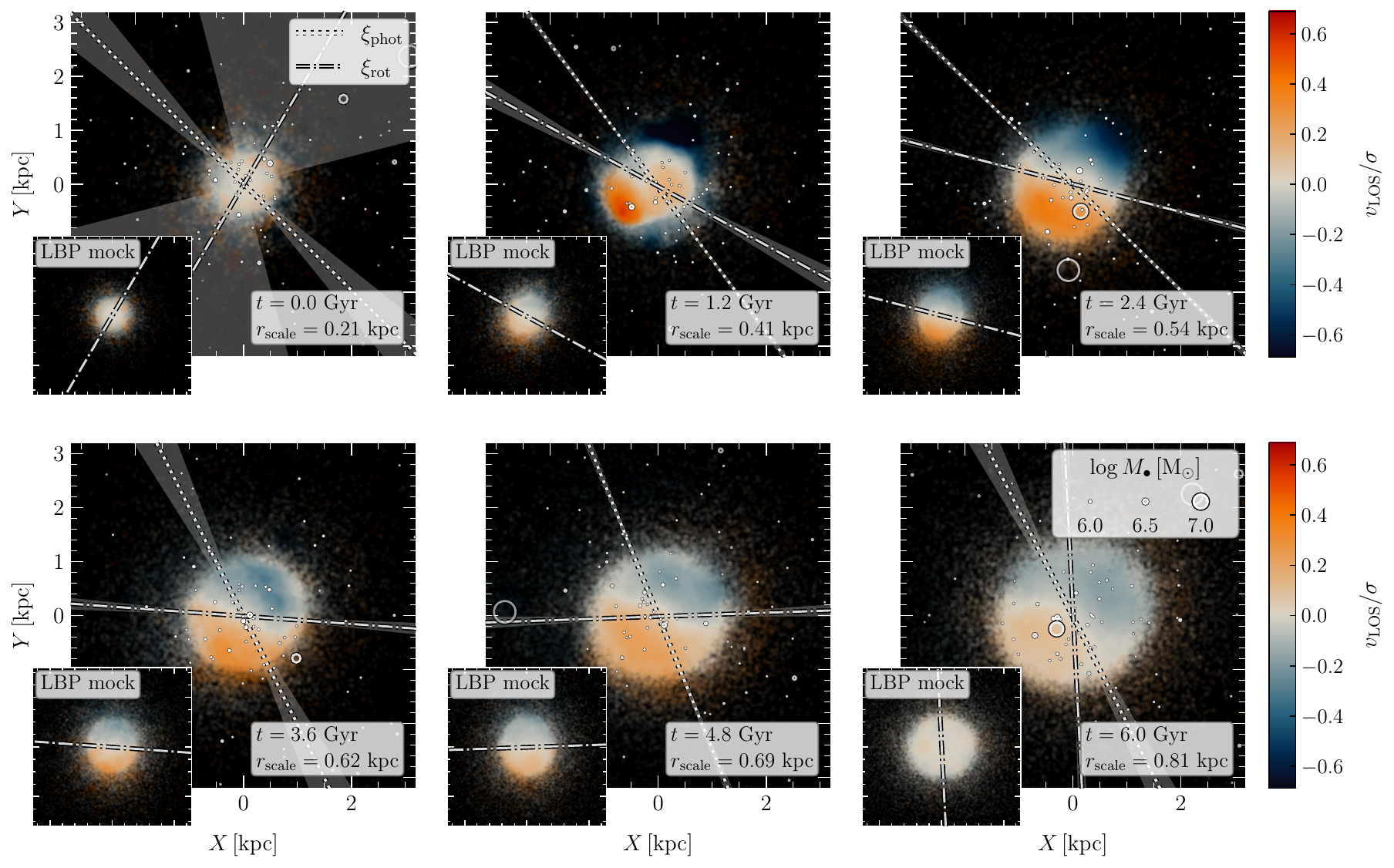}
\caption{\textit{Evolution of line-of-sight kinematics:}
Projected Cartesian maps showing the evolution of the normalised line-of-sight streaming motions of the simulated stars, $v_{\rm LOS}/\sigma$, for the model with parent dark matter halo mass $M_{\rm halo}=10^{9}~\msun$, with time increasing from left to right and from top to bottom.
Here, $\sigma \equiv \sigma_r$ is the spatially constant velocity-dispersion scale fitted independently at each snapshot.
In addition to the velocity structure, the panels also show the gradual expansion of the stellar spatial distribution, consistent with the gravothermal expansion identified in \protect\citet{Penarrubia+25}.
The lower-right text box in each panel gives the snapshot time, in Gyr, and the major-axis Plummer scale radius, $r_{\rm scale}$, in kpc.
All panels share the same dimensionless colour limits: blue denotes negative line-of-sight motion, red denotes positive line-of-sight motion, and white corresponds to zero streaming in the frame of the galaxy, after subtracting the fitted velocity zero point.
As in \protect\citet{Vitral+26}, the map transparency scales opposite to the local relative stellar surface density.
Subhaloes are shown as white circles, with radii proportional to their masses and transparencies increasing with distance from the centre of the dwarf.
The symmetry axis of the photometric distribution ($\xi_{\rm phot}$), obtained from the fits of \protect\citet{Vitral+26}, is shown as a dotted white line.
The symmetry axis of the fitted rotation model ($\xi_{\rm rot}$) is shown as a dot-dashed white line.
Both axes have a surrounding shaded region indicating their 16th--84th percentile uncertainty interval.
Each snapshot also includes, in the lower-left corner, an inset showing a mock realisation drawn from the best-fitting Lynden-Bell--Plummer velocity model, for which $\xi_{\rm phot} \equiv \xi_{\rm rot}$ by construction (cf. Appendix~\ref{ssec:rotation_mock_realisations}).
These inset panels share the same projected-coordinate and colour-bar limits as the main panels, allowing the simulated line-of-sight signatures to be compared directly with those expected from the best-fitting, physically motivated smooth rotating baseline.
At first glance, analyses based on line-of-sight kinematics alone would likely, and perhaps naively, interpret these snapshots as evidence for ordered rotation.
As shown below, however, the complementary velocity dimensions reveal departures from this simple interpretation.}
\label{fig:los-project-v-maps}
\end{figure*}

\begin{figure*}
\centering
\includegraphics[width=\hsize]{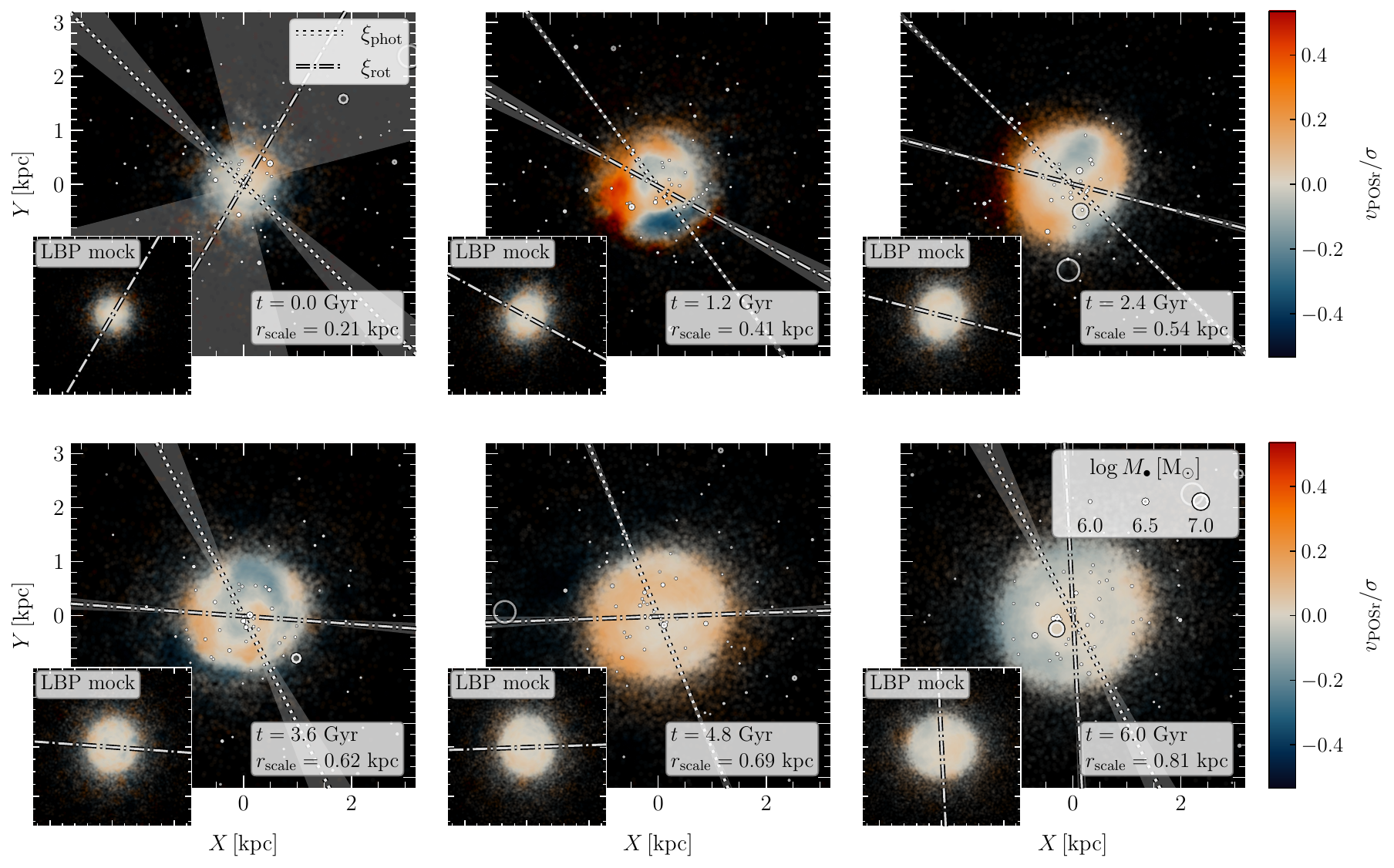}
\caption{\textit{Evolution of plane-of-sky radial kinematics:}
Same as Figure~\protect\ref{fig:los-project-v-maps}, but for the normalised plane-of-sky radial velocity component, $v_{\rm POSr}/\sigma$, using the same snapshot-by-snapshot normalisation as in that figure. In this projection, red denotes outward expansion of the stellar component, while blue denotes inward contraction. The maps show sporadic and irregular sign changes as a function of projected radius.}
\label{fig:posr-project-v-maps}
\end{figure*}

\begin{figure*}
\centering
\includegraphics[width=\hsize]{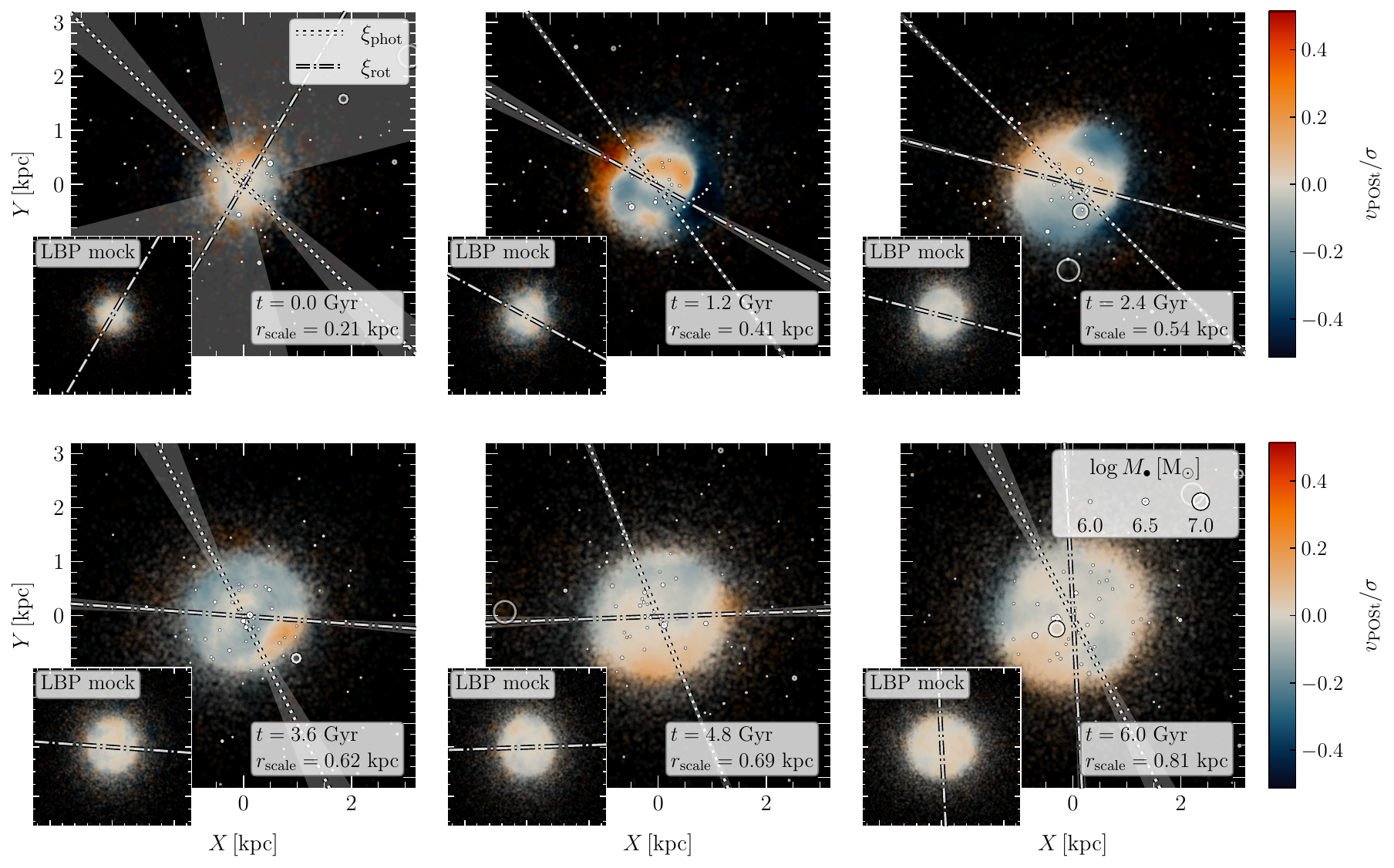}
\caption{\textit{Evolution of plane-of-sky tangential kinematics:}
Same as Figure~\protect\ref{fig:los-project-v-maps}, but for the normalised plane-of-sky tangential velocity component, $v_{\rm POSt}/\sigma$, using the same snapshot-by-snapshot normalisation as in that figure. In this projection, red denotes anticlockwise motion of the stellar component, while blue denotes clockwise motion. The maps show irregular sign changes as a function of projected radius and position angle, revealing alternating clockwise and anticlockwise streaming regions.}
\label{fig:post-project-v-maps}
\end{figure*}

\begin{figure*}
\centering
\includegraphics[width=\hsize]{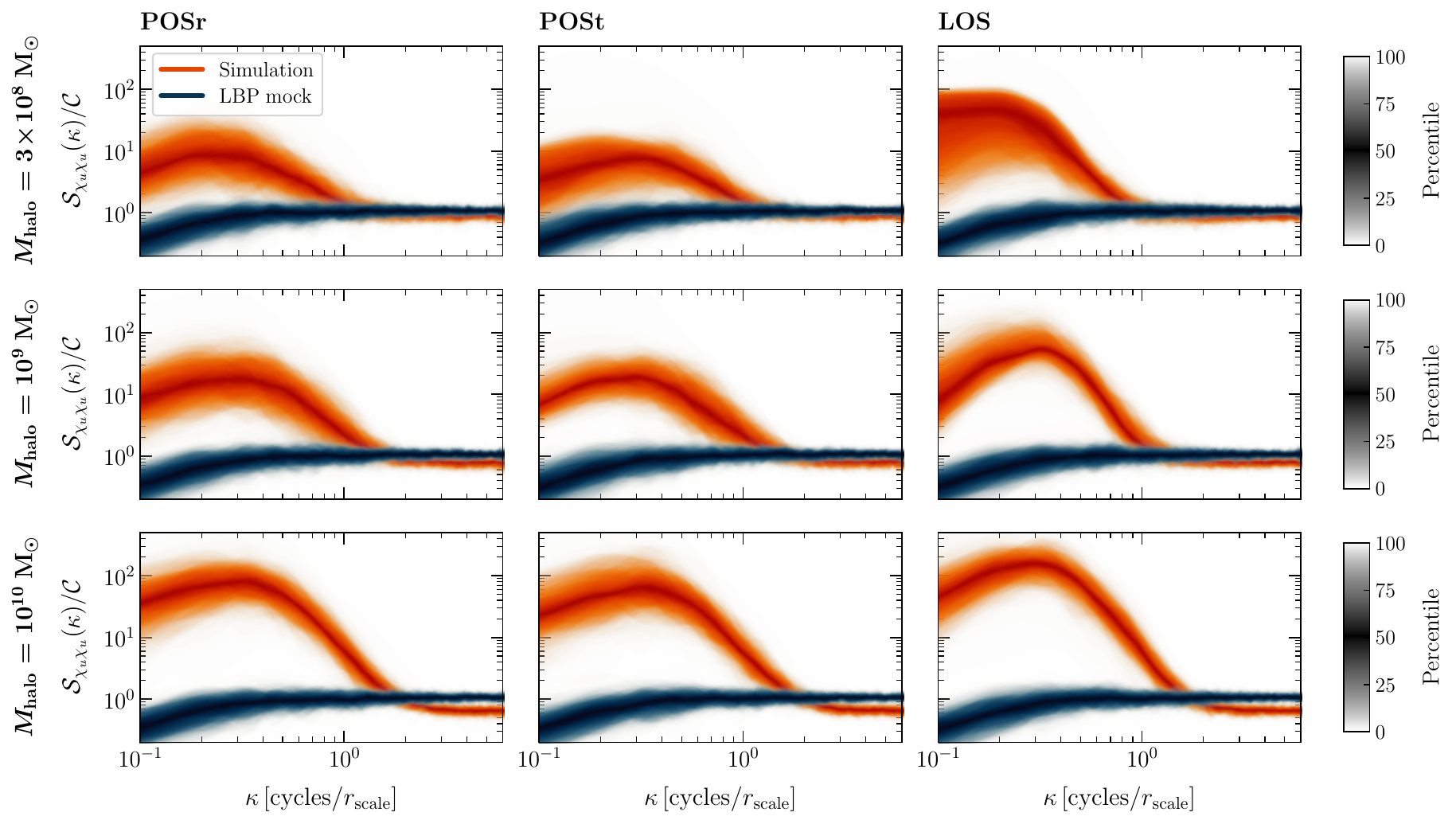}
\caption{\textit{Dependence on halo mass:}
Azimuthally averaged power-spectrum of the $\chi_{u}$ metric, defined in equation~\protect\ref{eq:methods_velocity_chi}, shown as the ratio $\mathcal{S}_{\chi_{u}\chi_{u}}/\mathcal{C}$, where $\mathcal{C}$ is the best-fitting noise floor. The horizontal axis gives the spatial frequency in units of cycles per $r_{\rm scale}$.
Columns show, from left to right, the plane-of-sky radial, plane-of-sky tangential, and line-of-sight velocity components. Rows show different parent dark matter halo masses, $M_{\rm halo} = {3 \times 10^{8},\,10^{9},\,10^{10}}~{\rm M}_{\odot}$, all for simulations with $N_{\star}=10^{5}$ stellar tracers. Each panel stacks the observable power-spectra from all snapshots in the corresponding simulation. The colour scale encodes percentile levels, with darker tones indicating more frequently occupied regions of the stacked distribution and lighter tones indicating less frequent ones. Red curves depict the observable for our subhalo experiments, while the blue ones show mock realisations of the best-fitting Lynden-Bell--Plummer model, as in previous figures. Although the more massive dark matter haloes source deeper potentials, the correspondingly more massive subhaloes implied by the common normalisation across halo-mass models produce more pronounced departures from the reference smooth kinematic structure. The line-of-sight component also consistently shows stronger power than the plane-of-sky components. Interestingly, for the lowest-mass halo, the signal appears more prominent in kinematic space than in the spatial analysis of Paper~I, as can be seen by comparison with the corresponding power-spectrum in their figure~5.}
\label{fig:ps-halo-mass}
\end{figure*}

\begin{figure*}
\centering
\includegraphics[width=\hsize]{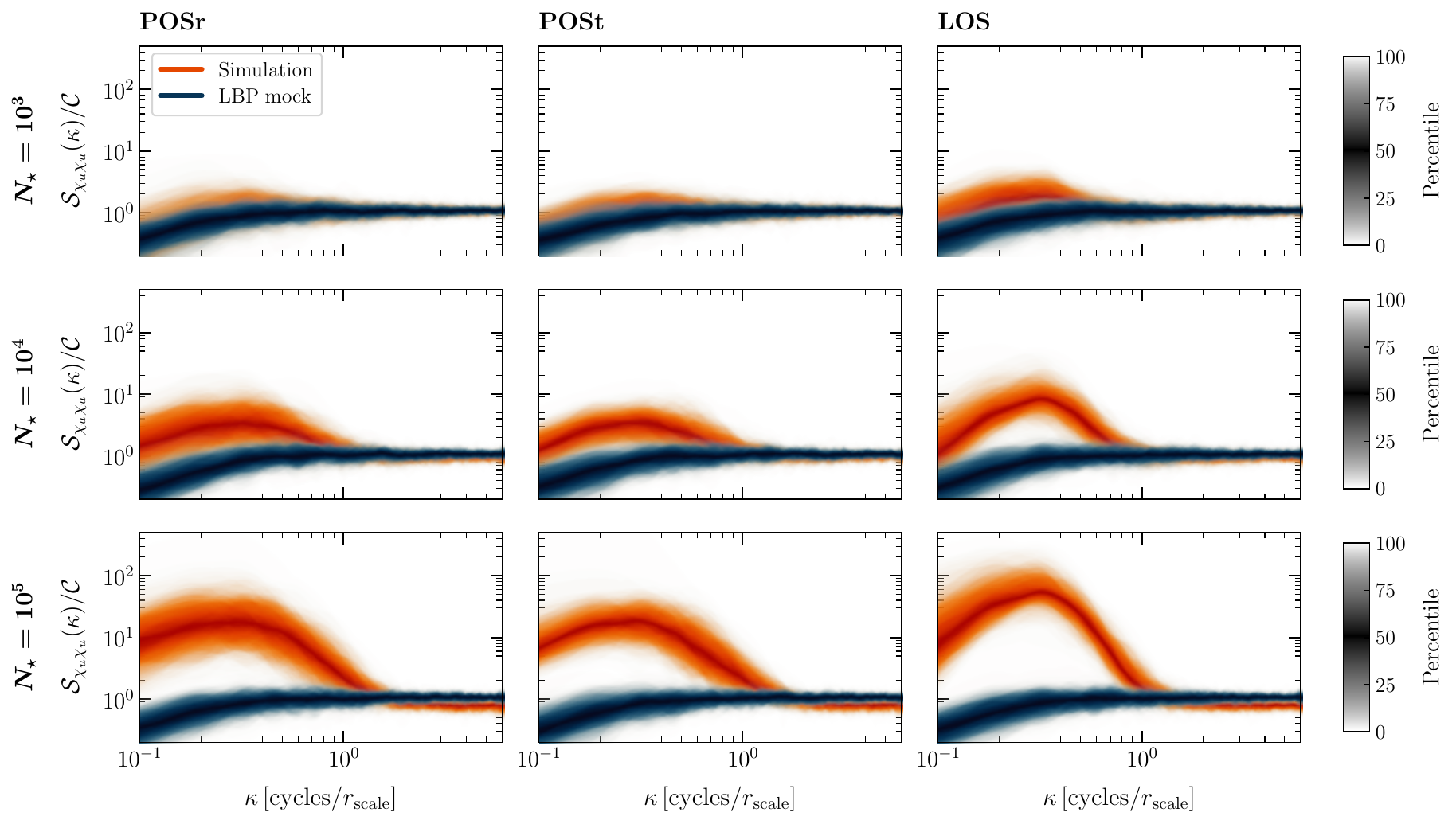}
\caption{\textit{Dependence on stellar counts:}
Same as Figure~\protect\ref{fig:ps-halo-mass}, but now fixing the parent halo mass to $M_{\rm halo} = 10^{9}~{\rm M}_{\odot}$ and varying the number of stellar tracers across $N_{\star} = \{10^{3},\,10^{4},\,10^{5}\}$. As expected, decreasing the number of tracers hampers the recovery of meaningful power-spectrum signatures. For $N_{\star}=10^{3}$, no robust departure from the reference smooth kinematic structure is detected, whereas for $N_{\star}=10^{4}$ some signal can already be recovered, especially in the line-of-sight component. The dependence on stellar counts therefore follows a trend similar to that found for the spatial features analysed in Paper~I.}
\label{fig:ps-stellar-count}
\end{figure*}

\subsection{Smooth rotating velocity model}
\label{ssec:methods_rotation_model}

At each simulation snapshot, the projected stellar density is described by the corresponding axisymmetric Plummer fit from Paper~I. This provides the centre, projected scale radius, axial ratio, and photometric position angle of the stellar distribution. We then fit the stellar velocities with a smooth rotating model motivated by the collisionless equilibrium family of \citet{LyndenBell67}. In this model, the ordered motion is purely azimuthal around an intrinsic symmetry axis, while the random motions follow associated Gaussian velocity moments.
The mean streaming velocity is parametrised as in equation~\ref{eq:rotation_law}, with the full mathematical formalism presented in Section~\ref{ssec:mathematical_formalism} of the appendix. In this configuration, the system behaves like a solid-body rotator at small radii, before the rotation profile declines in the outskirts. This choice is useful for the present problem because it provides a compact smooth baseline with enough flexibility to capture large-scale streaming motion, while retaining a well-defined distribution function. 
It also connects naturally to previous rotation models applied to stellar systems \citep{Bianchini+18_rotation_gc,Vasiliev19-rotation,Sollima+19,Arroyo-Polonio+24,Pascale+26}, many of which impose solid-body rotation. This corresponds to the limiting case of the present formalism in which the rotation scale radius, $r_{\phi}$, tends to infinity. Our fitted quantities can therefore be compared directly with related kinematic observables.

The same Lynden-Bell--Plummer model also fixes the relation between ordered and random motion. In particular, the maximum streaming velocity normalised by the constant radial velocity-dispersion scale, $(v/\sigma)$ (cf. equation~\ref{eq:lb_v_over_sigma}), is well defined for every fitted model. Larger values of this quantity therefore indicate a stronger contribution from coherent rotation relative to random motion.
 
The projected velocity moments are obtained by integrating the intrinsic Lynden-Bell--Plummer moments along the line of sight (cf. equation~\ref{eq:projected_mean_definition}). Importantly, the fit includes the projected kinematic symmetry axis angle, $\xi_{\rm rot}$. The kinematic axis is therefore not forced to coincide with the photometric position angle measured from the Plummer fit. This freedom is important because a mismatch between the two symmetry axes provides a simple way of testing whether the best-fitting smooth streaming field is aligned with the projected stellar distribution.

\subsection{Velocity residual fields}
\label{ssec:methods_velocity_residuals}

Once the smooth rotating model has been fitted, we compare the observed stellar velocities to the model prediction one star at a time. For a projected velocity component $u$, we define the normalised residual
\begin{equation}
    \chi_u
    =
    \frac{v_u-\langle v_u\rangle_{\rm model}}
    {\sigma_{u,{\rm model}}},
    \label{eq:methods_velocity_chi}
\end{equation}
where $\langle v_u\rangle_{\rm model}$ and $\sigma_{u,{\rm model}}$ are the projected mean velocity and dispersion predicted by the best-fitting smooth model at the position of the star. This normalisation places all velocity components on a common scale: values of $\chi_u$ measure departures in units of the local model dispersion.
For visualisation and diagnostics, we apply this construction to the projected Cartesian components, $v_{x'}$ and $v_{y'}$, to the local polar components on the plane of the sky (POS), $v_{\rm POSr}$ and $v_{\rm POSt}$, and to the line-of-sight (LOS) component, $v_{\rm LOS}$.

The resulting $\chi_u$ values define the velocity-space analogue of the density-fluctuation field used in Paper~I. We apply the same mapping procedure both to the simulation data and to forward mock realisations drawn from the fitted smooth model (see Section~\ref{ssec:rotation_mock_realisations} in the appendix for details). 
This allows residual structures in the simulations to be compared against the finite-sampling fluctuations expected from the baseline model itself. The resulting residual fields, and hence their power-spectra, remain tied to the adopted, physically motivated Lynden-Bell--Plummer model, and should therefore be interpreted as departures from that smooth rotating baseline.

\subsection{Fourier-space characterisation}
\label{ssec:methods_fourier}

As in Paper~I, we analyse the residual fields in Fourier space because this provides a natural way to separate noise-like fluctuations from coherent structure, while also reducing such structure to characteristic spatial scales that can be compared across halo models. Since these fields are sampled at the positions of individual stellar particles, rather than on a regular grid, we compute the Fourier amplitudes using a \textit{non-uniform} fast Fourier transform \citep*{Barnett+19,Barnett20}, as detailed in Section~\ref{ssec:nufft_power_spectra} of the appendix. Before applying the transform, we centre the projected coordinates on the Plummer fit, rotate them into the photometric-aligned frame, and express them in units of the instantaneous Plummer scale radius, $r_{\rm scale}$. Additionally, we subtract the mean value of the retained $\chi_u$ samples, preventing the zero-frequency mode from dominating the spectrum.

For each velocity component, we compute the two-dimensional power-spectrum of the residual field and then azimuthally average it as a function of the radial spatial frequency $\kappa$, in units of cycles/$r_{\rm scale}$. This produces a one-dimensional spectrum $\mathcal{S}_{\chi_u\chi_u}(\kappa)$ for each snapshot and velocity component. The azimuthal average removes phase information and therefore cannot describe all localised features in the velocity field, but it provides a compact summary of the characteristic spatial scales on which excess residual power is present. This makes the spectra easier to interpret, and compare across different models.

We process the simulation data and the fitted-model mocks in the same way. The mock spectra provide a baseline for the level and shape of residual power expected from a smooth Lynden-Bell--Plummer realisation with the same fitted parameters and sampling. As in the density-based analysis of Paper~I, this baseline is naturally shaped as a noise floor associated with finite-sampling fluctuations. Differences between the data and mock spectra therefore identify departures that are not absorbed by the smooth rotating model.

\subsection{Spectral model and summary diagnostics}
\label{ssec:methods_spectral_model}

To quantify the Fourier spectra, we fit each azimuthally averaged residual spectrum with two simple models. The null model contains only a constant high-frequency floor ($M_{0}$, equation~\ref{eq:spectral_model_m0}), while the one-feature model adds a single Voigt component ($M_{1}$, equation~\ref{eq:spectral_model_m1}).\footnote{This choice differs slightly from Paper~I, where two components were sometimes required; here, we find that one Voigt feature provides a satisfactory description of the residual power associated with subhalo-induced kinematic perturbations.}
The Voigt profile is flexible enough to represent both narrow and broad spectral features, while remaining simple enough to compare across different snapshots and simulation models.

The preferred model is selected independently for each spectrum by comparing $M_0$ and $M_1$ with an information criterion designed to penalise overfitting. We supplement this selection with basic resolution checks that reject unresolved or edge-truncated Voigt components (see Section~\ref{ssec:spectral_model_selection} for a full description). When $M_1$ is preferred, we characterise the detected feature through its fitted model parameters, in particular the central frequency, $\mu_{\kappa}$, used later in Section~\ref{ssec:power-spectrum-scale}.
If the null model is preferred, these feature-specific quantities are left undefined.

\section{Results} 
\label{sec:results}

The analysis of the kinematic components is substantially subtler than the treatment of the stellar spatial distribution presented in Paper~I. Perhaps the main reason for this is that, unlike the spatial distribution, the kinematics of dwarf galaxies affected by subhalo encounters do not evolve in a self-similar way, as already anticipated from figure~6 of P25. In that figure, apart from an overall scale factor, the spatial distribution remains well described by a Plummer profile at leading order, as also verified in figure~1 of Paper~I. By contrast, the evolution of the velocity-dispersion radial profile is marked by changes in both the inner and outer slopes, requiring a more flexible modelling strategy.

Here, we assume our physically motivated Lynden-Bell--Plummer models to provide a leading-order description of the smooth rotating kinematic field.
However, a first caveat must be acknowledged: although physically motivated (see beginning of Section~\ref{ssec:mathematical_formalism}), this is by no means the only plausible distribution function capable of describing the leading-order dynamics of a dwarf galaxy. 
Nonetheless, as we shall show later in Section~\ref{ssec:rotate_not-rotate}, there are specific departures from this model that may indicate transient and out-of-equilibrium signatures in a sense broader than this particular family of kinematic models.
The following subsections thus aim to develop a broader understanding of such signatures, their dependence on modelling choices, and the general features that could be associated with the presence of a subhalo population in dwarf spheroidal galaxies.

\subsection{Projected kinematic maps}
\label{ssec:kinematic-maps}

To start the presentation of our results, we first place the analysis in the context of typical searches for rotation signatures in dwarf spheroidal galaxies, and more generally in stellar systems. Until the latest \gaia\ data releases, the only kinematic component available with sufficient precision to model rotation in such systems was the line-of-sight velocity. Even with \hst's superb astrometric capabilities, the usual treatment of systematics in internal kinematic analyses tends to erase first-order moments in the proper motion fields of Local Group stellar systems \citep{Bellini+14}.\footnote{\hst\ raw proper motions may contain low-level systematic effects associated with charge-transfer-efficiency degradation in the \hst\ charge-coupled devices, as well as with small epoch-to-epoch variations in the geometric-distortion solution. These effects can produce spatially coherent proper-motion offsets across the observed fields. They are commonly mitigated through local corrections, but such corrections effectively suppress genuine streaming signals globally.}
This has largely limited \hst-based studies of Milky Way satellites to pressure- and dispersion-based analyses \citep[e.g.][]{Watkins+15,Vitral+24}, rather than direct measurements of their intrinsic streaming motions.

Figure~\ref{fig:los-project-v-maps} shows the time evolution of the normalised line-of-sight velocity field of the simulated galaxy.\footnote{For readability, we first use these maps to introduce the main phenomenology of the kinematic fields. The technical details of their construction are described in Section~\ref{ssec:power-spectrum-scale}. Importantly, we also verified that the main conclusions are unchanged when adopting different projection frames.}
Here, $\sigma \equiv \sigma_r$ is the spatially constant velocity-dispersion scale fitted independently at each snapshot (see equation~\ref{eq:sigma-definition}). Alongside the changing velocity structure, the panels also display the gradual expansion of the stellar distribution, consistent with the gravothermal expansion identified by P25. The shared colour scale is defined in the galaxy frame, after subtracting the fitted velocity zero point: blue denotes blueshifted motion, red denotes redshifted motion, and white corresponds to the stationary state. As in Paper~I, the map opacity increases with the local relative stellar surface density. Subhaloes are shown as white circles, with radii proportional to their masses and transparencies increasing with distance from the dwarf centre. For reference, the photometric symmetry axis from the fits of Paper~I is shown as a dotted white line ($\xi_{\rm phot}$), while the symmetry axis of the fitted rotation model is shown as a dot-dashed white line ($\xi_{\rm rot}$). The shaded regions around them mark the 16th--84th percentile uncertainty interval, which is naturally broad in the first snapshot because its streaming signal is negligible at this early time. 
For each snapshot, the lower-left inset shows a forward mock realisation of the best-fitting, physically motivated smooth rotating baseline, enabling a direct comparison with the simulated velocity field.\footnote{We note that, in these mock realisations, $\xi_{\rm phot} \equiv \xi_{\rm rot}$ by construction.}

Taken at face value, the line-of-sight maps alone resemble the kind of rotation pattern commonly searched for in stellar systems. However, this interpretation becomes less straightforward once the complementary kinematic dimensions are considered, in particular the radial and tangential components of the plane-of-sky projected velocity field (Figures~\ref{fig:posr-project-v-maps} and \ref{fig:post-project-v-maps}, respectively). Both the $v_{\rm POSr}$ and $v_{\rm POSt}$ components often change sign as a function of projected radius and position angle. For $v_{\rm POSr}$, this indicates multiple zones of contraction and expansion relative to the galaxy centre, whereas for $v_{\rm POSt}$ it reveals streaming motions in both clockwise and anticlockwise directions across the galaxy.
As described in Section~\ref{ssec:power-spectrum-scale}, these maps are smoothed on scales selected from the excess power of the residual velocity field, after being normalised by the local velocity dispersion. The resulting structures therefore do not simply reflect random stellar motions around a smooth baseline, but rather coherent velocity shifts relative to it.
This distinction will be important in Section~\ref{ssec:rotate_not-rotate}, where we argue that these maps should be interpreted not only as departures from the adopted Lynden-Bell--Plummer kinematic baseline, but as broader departures from the sign symmetries expected for stationary rotation in an axisymmetric stellar system.

\subsection{Characteristics of the power-spectrum}
\label{ssec:power-spectrum-results}

\subsubsection{Relevant spatial scales}
\label{ssec:power-spectrum-scale}

The previous subsection showed that the projected velocity maps of our simulations display intricate and irregular sign changes in the plane-of-sky radial and tangential components, while still preserving a line-of-sight signature somewhat reminiscent of the rotation patterns searched for in observational campaigns. A first point to clarify is therefore how these maps were constructed so that such signatures can be identified robustly. At a basic level, the construction is straightforward: for each star, we consider the set of neighbours within a projected radius $R_{\lambda}$ and compute the median velocity in the analysed component. However, the choice of $R_{\lambda}$ is important. If this scale is too small, the maps may amplify stochastic fluctuations; if it is too large, genuine localised signatures may be smoothed away. The relevant question is therefore how to choose $R_{\lambda}$ so as to recover the signatures of interest without introducing an arbitrary smoothing scale.

To address this, we use the power-spectrum of the $\chi_{u}$ measure defined in equation~\ref{eq:methods_velocity_chi}. This quantity is designed to down-weight random variations associated with the intrinsic velocity dispersion of the system, while emphasising signatures that depart from a smooth kinematic baseline, here taken to be the Lynden-Bell--Plummer family formulated in Appendix~\ref{ssec:mathematical_formalism}. The power-spectrum of $\chi_u$ then indicates the spatial frequencies at which these departures are strongest, through an excess that is well described by a Voigt component. When constructing the velocity maps, we therefore choose $R_{\lambda}$ at each snapshot from the peak frequency $\mu_{\kappa}$ of the best-fitting Voigt component in equation~\ref{eq:spectral_model_m1}, using
\begin{equation}
R_{\lambda} = \frac{1}{2\pi\mu_{\kappa}}.
\label{eq:smoothing-length}
\end{equation}
This choice reflects the fact that $\mu_{\kappa}$ is expressed in cycles per unit length: rather than averaging over a full cycle, we probe the characteristic substructure captured throughout the cycle itself.\footnote{Because there are three velocity components, we use the same smoothing scale for all of them. In practice, we select the component that maximises $\log_{10}(f/\mathcal{C})+\log_{10}\mu_{\kappa}$, where $(f/\mathcal{C})$ is the normalised spectral flux defined in Appendix~\ref{ssec:spectral_model_selection}, such that we penalise both very low $\mu_{\kappa}$ values and weak signals. This choice makes little practical difference, since the three $\chi_u$ fields are associated with broadly similar characteristic spatial scales.} In this way, $R_{\lambda}$ is not chosen arbitrarily, but is instead tied to the characteristic scale at which the $\chi_u$ field contains the strongest fitted excess power. This allows the diagnostics shown in Figures~\ref{fig:los-project-v-maps}, \ref{fig:posr-project-v-maps}, and \ref{fig:post-project-v-maps} to highlight coherent velocity departures without being dominated by either stochastic fluctuations or excessive smoothing.

\begin{figure*}
\centering
\includegraphics[width=\hsize]{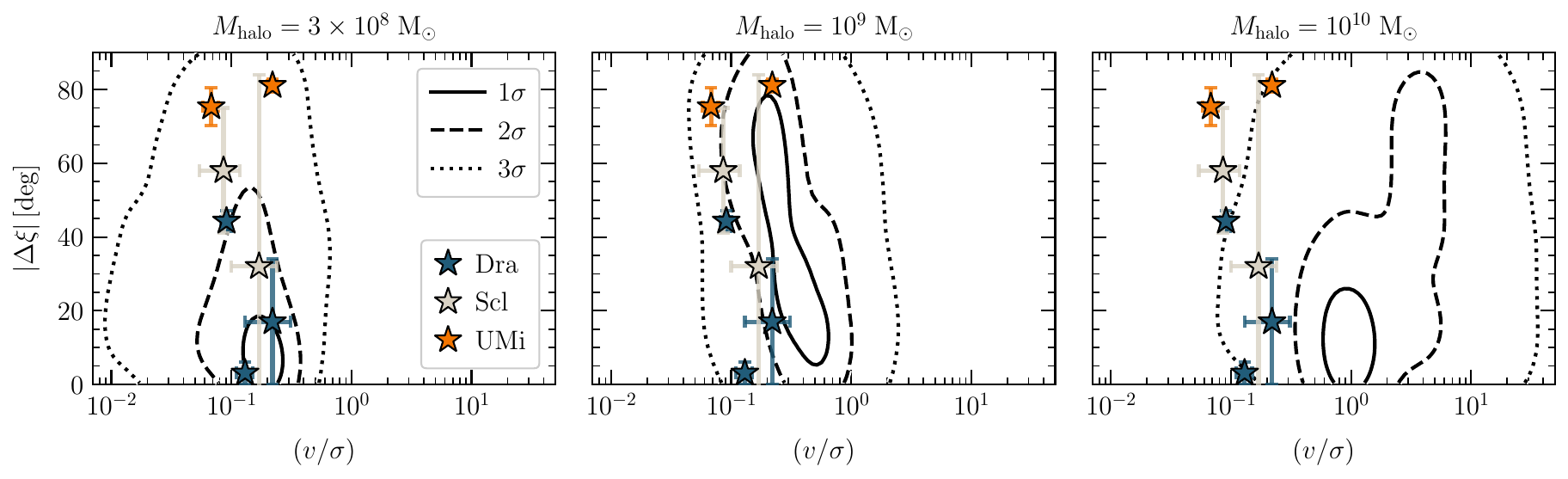}
\caption{\textit{Rotation metrics:}
Snapshot density of the absolute projected misalignment between the kinematic and photometric axes, $|\Delta \xi|$, versus the rotation ratio $(v/\sigma)$, as defined in equation~\ref{eq:lb_v_over_sigma} of the Appendix. The misalignment ranges from $0$ degrees for oblate-like rotation to $90$ degrees for prolate-like rotation. For each simulation snapshot, the measured position and its uncertainties are represented by sampling the corresponding asymmetric error distributions, thereby convolving the distribution of snapshots with their measurement errors. Black contours trace the sampled-point density at 2D-Gaussian-equivalent $1$, $2$, and $3\sigma$ heights relative to its maximum. Coloured stars show recent observational estimates for classical Milky Way dwarf spheroidals: Draco in blue, from \protect\citet{Vitral+24} using the data of \protect\citet{Walker+23}, and from \protect\citet{Pascale+26} using data of \protect\citet{Walker+23} and \protect\citet{Geha+26}, separately; Sculptor in grey, from \protect\citet{Vitral+26-sculptor} using the combined data of \protect\citet{Walker+23} and \protect\citet{Tolstoy+23}, and from \protect\citeauthor{Arroyo-Polonio+24} (\protect\citeyear{Arroyo-Polonio+24}, evaluated at the half-light radius) using data of \protect\citet{Tolstoy+23}; and Ursa Minor in orange, from \protect\citet{Pascale+26} using data of \protect\citet{Walker+23} and \protect\citet{Geha+26}, separately. The simulated distributions depend on the initial halo-mass and subhalo orbital configuration: in the model with the least massive subhaloes, the induced misalignments remain smaller and $(v/\sigma)$ concentrates around $\gtrsim 0.1$, whereas the models with more massive subhaloes span a broad range of $|\Delta \xi|$, covering the full allowed interval. Their typical $(v/\sigma)$ values are denser around $\gtrsim 0.2$ for the $M_{\rm halo}=10^{9}~{\rm M}_{\odot}$ model and around $\gtrsim 1$ for the $M_{\rm halo}=10^{10}~{\rm M}_{\odot}$ model. Our experiments therefore populate the range of $|\Delta \xi|$ and $(v/\sigma)$ values inferred in recent measurements of classical Milky Way dwarf spheroidals.}
\label{fig:delta-xi_v-sigma}
\end{figure*}

\subsubsection{Dependence on halo mass and stellar counts}
\label{sssec:power-spectrum-halo-counts}

As in Paper~I, it is useful to assess how the results presented so far, and in particular their power-spectrum description, vary with both the underlying dark matter potential and the number of stellar tracers used in the analysis. Following that work, we therefore test the robustness of the observed signatures across different halo-mass models, $M_{\rm halo} = \{3 \times 10^{8},\,10^{9},\,10^{10}\}~{\rm M}_{\odot}$, in Figure~\ref{fig:ps-halo-mass}, and across different stellar counts, $N_{\star} = \{10^{3},\,10^{4},\,10^{5}\}$, in Figure~\ref{fig:ps-stellar-count}. In both cases, we show the floor-normalised azimuthally averaged power-spectrum, $\mathcal{S}_{\chi_{u}\chi_{u}}/\mathcal{C}$, as a function of spatial frequency, with columns corresponding to the plane-of-sky radial, plane-of-sky tangential, and line-of-sight velocity components. The panels stack the observable spectra across all snapshots in each simulation, with darker percentile regions indicating more frequently occupied parts of the distribution. The red curves show this observable for our subhalo experiments, while blue curves show the corresponding kinematic reference realisations drawn from the best-fitting Lynden-Bell--Plummer model.

Figure~\ref{fig:ps-halo-mass} shows that the spectral departures from the reference smooth kinematic structure become more pronounced for increasing parent halo mass. Although the more massive haloes source deeper and more stable potentials, the common normalisation of the subhalo population implies correspondingly more massive perturbers, leading to stronger kinematic signatures (see Figure~\ref{fig:subhalo-masses}). Across all halo masses, the line-of-sight component consistently exhibits stronger power than the two plane-of-sky components. Interestingly, even the lowest-mass halo produces a visibly relevant kinematic signal, which appears more prominent than the corresponding spatial signal discussed in Paper~I (see their Figure~5, respectively). Figure~\ref{fig:ps-stellar-count} then shows that decreasing the number of tracers progressively hampers the recovery of these signatures. For $N_{\star}=10^{3}$, no robust departure is detected, whereas for $N_{\star}=10^{4}$ some signal can already be recovered, especially in the line-of-sight component. The dependence on stellar counts therefore follows a trend similar to that found for the spatial features in Paper~I, while confirming that the kinematic power-spectrum remains informative once the tracer sample is sufficiently large.

\subsection{Rotation metrics and kinematic-axis misalignment}
\label{ssec:delta-xi_v-over-sigma}

Measurements of rotation in dwarf spheroidal galaxies remain scarce compared to globular clusters \citep{Bianchini+18_rotation_gc,Sollima+19,Vasiliev19-rotation} or dwarf and giant ellipticals \citep{Franx&Illingworth98,DeRijcke+04,Howley+13}. Nevertheless, robust line-of-sight samples with $\gtrsim \mathcal{O}\left(10^{2}\right)$ stellar tracers have now enabled rotation measurements in a handful of classical Milky Way satellites, including Ursa Minor \citep{Pace+20,Pascale+26}, Draco \citep{Vitral+24,Pascale+26}, and Sculptor \citep{Battaglia+08,Arroyo-Polonio+24,Vitral+26-sculptor}. These studies are particularly relevant here because their models allowed the projected kinematic axis to be misaligned with the photometric axis, although earlier works had already searched for rotation using more restricted prescriptions \citep{Walker+08,Strigari10,Zhu+16}.
The reported rotation signals are generally weak, with $(v/\sigma)\sim \mathcal{O}\left(10^{-1}\right)$, and they usually span a broad range of projected configurations between the oblate-like and prolate-like limits.
The reader can find representative examples for Draco in figure~4 of \citet{Vitral+24} and figure~6 of \citet{Pascale+26}; for Sculptor in figure~6 of \citet{Arroyo-Polonio+24} and figures~4 and~8 of \citet{Vitral+26-sculptor}; and for Ursa Minor in figure~9 of \citet{Pace+20} and figure~7 of \citet{Pascale+26}.

This observational context motivates a direct comparison with our controlled simulations, which are initialised without intrinsic rotation. We therefore consider two projected rotation diagnostics: the rotation ratio $(v/\sigma)$, defined in equation~\ref{eq:lb_v_over_sigma}, and the absolute misalignment between the projected kinematic and photometric axes, $|\Delta \xi|$. The latter ranges from $0$ degrees for oblate-like rotation to $90$ degrees for prolate-like rotation. As discussed in Appendix~\ref{ssec:rotation_recovery_tests}, $(v/\sigma)$ remains robustly recovered even for samples with $10^{3}$ stellar tracers, while $|\Delta \xi|$ can also be recovered, albeit with substantially larger uncertainties. These two quantities therefore provide a compact way of comparing the simulations with current observational measurements.
Figure~\ref{fig:delta-xi_v-sigma} shows the resulting uncertainty-convolved snapshot density contour in the $|\Delta \xi| - (v/\sigma)$ plane for the three halo-mass models considered in this work. We deliberately emphasise the density of sampled snapshots rather than a single time-ordered trajectory, because the detailed evolution depends sensitively on the orbital configuration of the subhalo population. In particular, individual massive passages can perturb the stellar component at different times and directions, so that the same halo-mass model need not follow a unique monotonic track in this plane.

The contours nevertheless reveal a clear statistical dependence on the halo--subhalo mass scale. The model with the least massive subhaloes preferentially occupies low rotation ratios, with $(v/\sigma)$ concentrated around $\gtrsim 0.1$, and comparatively small misalignments. The intermediate-mass model, with $M_{\rm halo}=10^{9}~{\rm M}_{\odot}$, spans a wide range of $|\Delta \xi|$ and has its densest region around $(v/\sigma)\gtrsim 0.2$, overlapping well with the values inferred for the classical dwarf spheroidals shown in the figure. The most massive halo model also spans a broad range of projected misalignments, but its rotation ratios are typically larger, with substantial density around $(v/\sigma)\gtrsim 1$ and a tail extending to still higher values. This latter regime is less representative of the observed systems considered here, and is also less directly comparable to them if interpreted literally, since the upper end of its subhalo mass spectrum, as shown in Figure~\ref{fig:subhalo-masses}, can approach the halo masses inferred for classical dwarfs themselves \citep[e.g.][]{Vitral+24,Vitral+26-sculptor}.

To check that this interpretation is not driven solely by the particular stochastic realisation used in the main figures, we also analysed ten additional realisations of each halo-mass model at a fixed late-time snapshot.\footnote{We restrict this stochastic-realisation test to a single late-time snapshot per halo-mass model because the full rotation-fitting procedure is computationally expensive, even with parallel execution on high-performance computing facilities. Extending the same analysis to all snapshots and respective multiple realisations would require a substantially larger dedicated numerical campaign.}
These tests confirm that $(v/\sigma)$ increases with halo mass only in a statistical sense: the typical values shift upward for more massive subhalo populations, but the distributions are broad and outliers are expected. The same tests show that $|\Delta \xi|$ remains widely distributed, with no simple monotonic dependence on halo mass. The scatter recovered from these additional realisations is broadly consistent with the regions enclosed by the snapshot-density contours in Figure~\ref{fig:delta-xi_v-sigma}, supporting the use of those contours as a compact summary of the stochasticity associated with subhalo orbital configurations.

Taken together, the comparison suggests that models with $M_{\rm halo}\lesssim 10^{9}~{\rm M}_{\odot}$ can populate the region of projected rotation signal and axis misalignment currently measured in classical Milky Way dwarf spheroidals, despite being initialised without intrinsic rotation. However, this agreement should not be interpreted as a unique identification of subhalo-driven rotation: the simulations are intentionally idealised, and additional effects such as tides, baryons, mergers, and more general equilibrium configurations may also affect the same observables. We return to these caveats in Section~\ref{ssec:alternative-perturbations}, where we discuss how the residual velocity maps and power-spectra may help distinguish between different perturbative mechanisms.

\begin{figure*}
\centering
\includegraphics[width=\hsize]{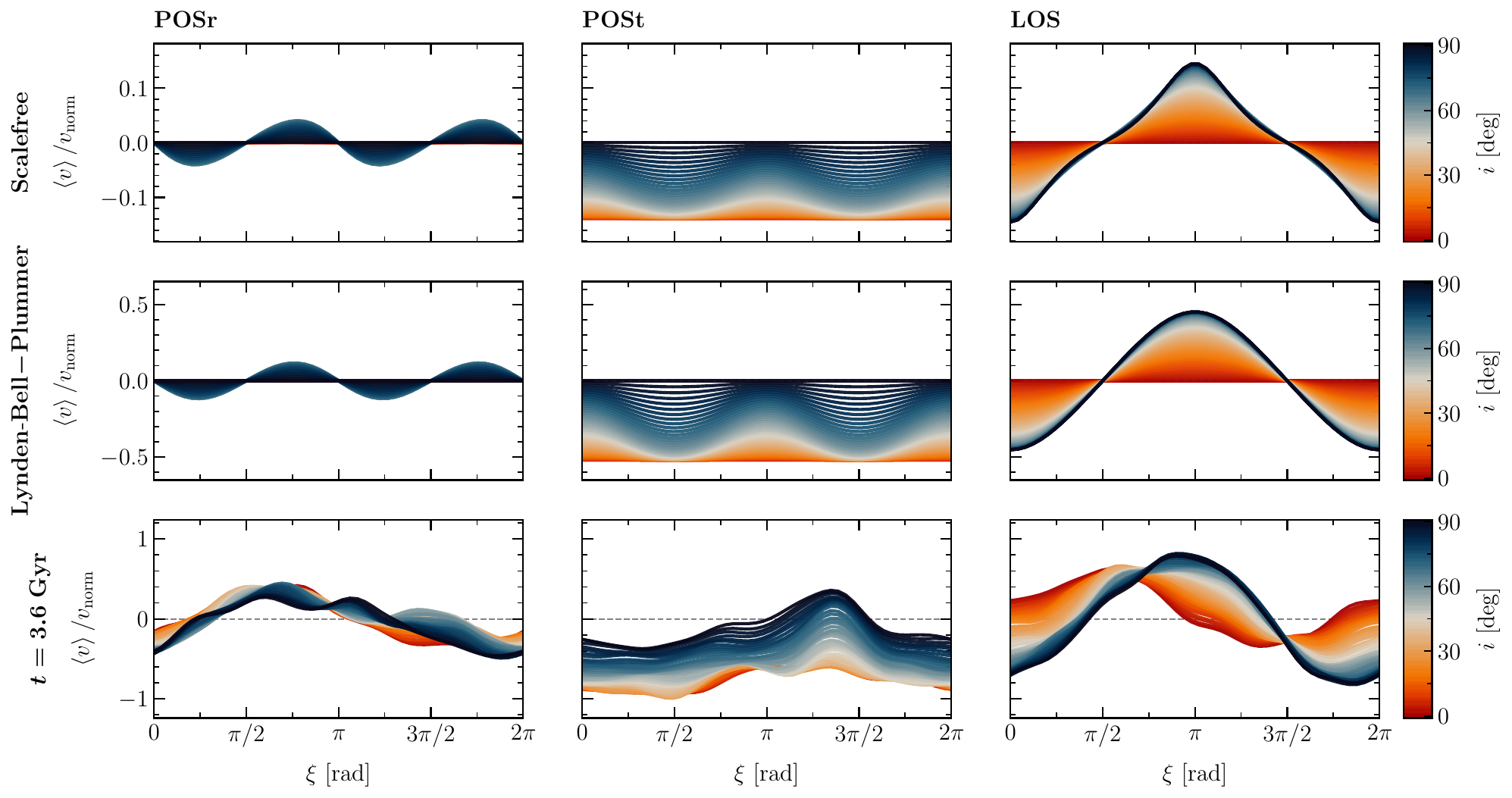}
\caption{\textit{Projected signatures of stationary axisymmetric rotation:}
Normalised projected first velocity moments, $\langle v\rangle/v_{\rm norm}$, as a function of position angle $\xi$ for two smooth rotating reference families and one simulated snapshot. Columns show, from left to right, the plane-of-sky radial, plane-of-sky tangential, and line-of-sight velocity components, with $\langle v\rangle/v_{\rm norm}=0$ marked by a dotted line for reference. The first row shows a fiducial scale-free axisymmetric model from the family of \protect\citet{deBruijne+96}, using the parameter choices described in the main text; these choices affect the detailed amplitudes and curve shapes, but not the conclusions drawn below. The second row shows the corresponding projected first moments for the Lynden-Bell--Plummer kinematic baseline adopted in this work, evaluated at $R_{\rm POS}=r_{\phi}/2$. The third row shows a representative snapshot of the $M_{\rm halo}=10^{9}~\msun$ simulation, evaluated at the same relative radius and Gaussian-smoothed with an angular width $R_{\lambda}/R_{\rm POS}$, where $R_{\lambda}$ is defined in equation~\ref{eq:smoothing-length}. For visual comparison, $\xi$ in this row is measured from the fitted kinematic major axis rather than from the photometric major axis. Although the snapshot has a single fitted projected inclination, we reproject it over the same inclination range as the reference models in the upper rows in order to illustrate the viewing-angle dependence of the observables. Curves are coloured by inclination, from face-on ($i=0$ degrees) to edge-on ($i=90$ degrees). In the first two rows, the relevant feature is not the precise amplitude of the curves, but the regular sign structure expected for stationary axisymmetric rotation: $v_{\rm POSr}$ and $v_{\rm LOS}$ follow ordered quadrant sequences, while $v_{\rm POSt}$ does not alternate sign around the projected body. By contrast, the representative simulation snapshot shows departures from these sinusoidal symmetries.}
\label{fig:rotation-profiles}
\end{figure*}

\begin{figure*}
\centering
\includegraphics[width=\hsize]{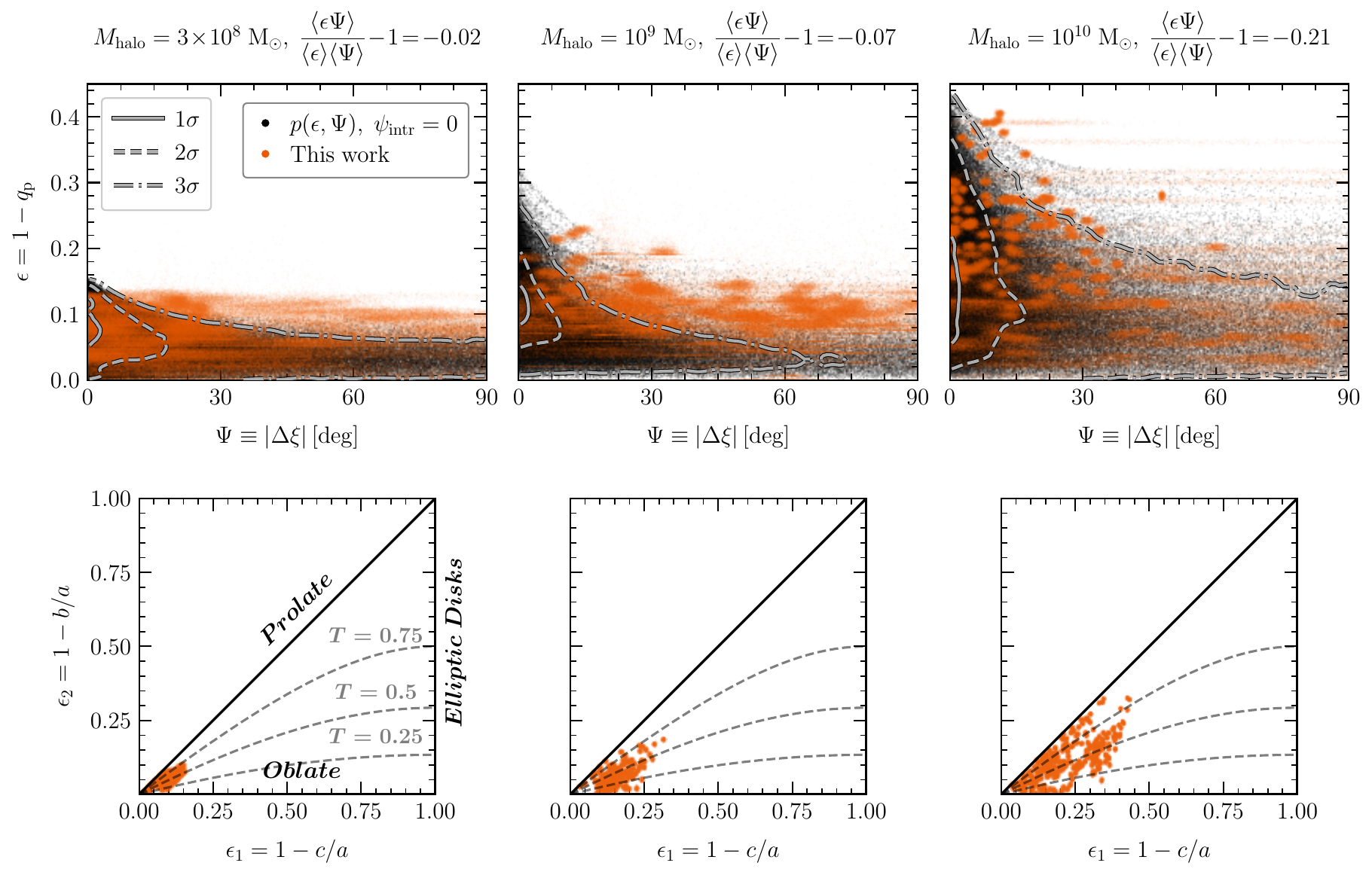}
\caption{\textit{Kinematic misalignment:}
\textbf{Upper row --} projected ellipticity, $\epsilon$, versus the projected
misalignment between the photometric and kinematic symmetry axes,
$\Psi\equiv|\Delta\xi|$. Black points show the expected distribution of
$(\epsilon,\Psi)$ for triaxial systems with no \textit{intrinsic}
misalignment ($\psi_{\rm int}=0$). For each snapshot, we sample its inferred intrinsic-shape posteriors and project the resulting shapes over isotropic viewing directions, before combining all snapshots within a given halo-mass model. The grey curves mark the corresponding $1$, $2$, and $3\sigma$ 2D Gaussian-equivalent iso-height density contours. Orange points show the measurements from our subhalo experiments, convolved with their asymmetric uncertainties. The panel titles additionally report the correlation statistic introduced by \protect\citet{Franx+91},
$\langle\epsilon\Psi\rangle/(\langle\epsilon\rangle\langle\Psi\rangle)-1$, for the ensemble of snapshots from our subhalo experiments.
\textbf{Lower row --} corresponding distributions of the inferred intrinsic shape
parameters, $\epsilon_1=1-c/a$ and $\epsilon_2=1-b/a$ (see Equations~\ref{eq:triaxial_intrinsic_eps} and \ref{eq:triaxial_T}). The boundaries indicate the oblate, prolate, and infinitely thin-disk limits, while the intermediate curves trace constant triaxiality $T$. Increasing halo mass shifts the stellar distribution away from near-sphericity towards more elongated shapes, spanning triaxial, oblate- and prolate-like configurations alike.
The two lower-mass models show no significant $\epsilon$--$\Psi$ correlation and exhibit an excess of systems with simultaneously high $\epsilon$ and $\Psi$ relative to the intrinsically aligned prediction. By contrast, the $10^{10}~\msun$ model yields
$\langle\epsilon\Psi\rangle/(\langle\epsilon\rangle\langle\Psi\rangle)-1=-0.21$ and its full $(\epsilon,\Psi)$ distribution agrees remarkably well with the aligned expectation. Overall, the lower-mass models therefore indicate that their projected photometric--kinematic misalignments cannot be readily explained by projection effects alone, suggesting the development of a non-zero intrinsic misalignment, $\psi_{\rm int}\neq0$.}
\label{fig:triaxiality-misalignment}
\end{figure*}

\begin{figure}
\centering
\includegraphics[width=\hsize]{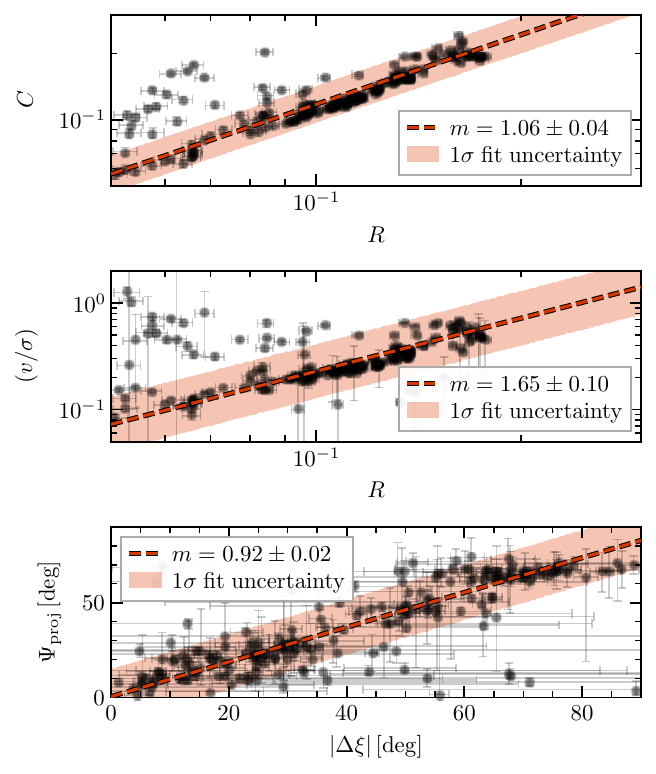}
\caption{\textit{Three-dimensional streaming coherence and projected rotation
diagnostics:}
Correlations between the intrinsic angular-momentum coherence of the stellar
component and the quantities inferred from our projected
Lynden-Bell--Plummer fits, shown for the
$M_{\rm halo}=10^{9}~{\rm M}_{\odot}$ model. From top to bottom, we show
$C$ versus $R$ (cf. Equation~\ref{eq:Cj_definition}), $(v/\sigma)$ versus $R$, and the projected
misalignment $\Psi_{\rm proj}$ between the sky projections of the intrinsic
triaxial short axis and the preferred three-dimensional angular-momentum
axis versus the independently fitted photometric--kinematic misalignment
$|\Delta\xi|$. Black points show individual simulation snapshots with their
corresponding uncertainties. Dashed lines show the
orthogonal-distance-regression fits, while the shaded regions show the
pointwise 16th--84th percentile prediction intervals obtained by sampling
the joint uncertainty in the fitted slope and intercept and including the
empirical scatter of the snapshots about the fitted relation. The fitted
slope and its $1\sigma$ uncertainty are reported in each panel. Only
snapshots satisfying the adopted lower cuts $C>0.05$, $R>0.05$, and
$(v/\sigma)>0.05$ are included in the comparison, in order to limit the
influence of weak-signal measurements for which noise and finite-sampling
effects can disproportionately affect the inferred coherence, rotation
amplitude, and preferred streaming direction.}
\label{fig:correlations-streaming}
\end{figure}

\section{Discussion} \label{sec:discussion}

\subsection{To rotate or not to rotate?}
\label{ssec:rotate_not-rotate}

To start the discussion of our results, we ask the following broader question: should the streaming motions shown in Figures~\ref{fig:los-project-v-maps}, \ref{fig:posr-project-v-maps}, and \ref{fig:post-project-v-maps} be understood as rotation in the usual stationary sense, or as a different form of transient streaming? The line-of-sight maps alone could naturally suggest the former. They often resemble the blueshift--redshift pattern commonly used to identify rotation in stellar systems, and our fits indeed recover non-negligible $(v/\sigma)$ values for most snapshots. The plane-of-sky components, however, complicate this interpretation. The $v_{\rm POSr}$ maps show alternating patches of expansion and contraction, while the $v_{\rm POSt}$ maps show regions of clockwise and anticlockwise streaming across the stellar body. This suggests that the measured rotation-like signal is only a leading-order projection of a more complex velocity field.

\subsubsection{Axisymmetric expectations}
\label{sssec:axisymmetry}

To make this concept more concrete, it is useful to recall what stationary axisymmetric rotation is expected to look like in projection. For this purpose, we use the scale-free distribution-function models of \citet*{deBruijne+96}, implemented in the publicly available \scf\ code.\footnote{Code repository:\\ \url{https://gitlab.com/eduardo-vitral/scalefree/}} These models describe axisymmetric stellar systems embedded in spherical potentials and provide both intrinsic and projected velocity moments. Their scale-free nature is particularly useful here: because the velocity moments are normalised and not tied to a fixed physical radius, the resulting angular patterns can be interpreted as generic signatures of stationary rotation at any radius. The original formalism treats the line-of-sight kinematics, while the plane-of-sky extension follows Appendix~B of \citet{Vitral+24}. Although this family of models is by no means exhaustive, it offers a physically motivated reference with which to build intuition for the projected sign symmetries expected from stationary axisymmetric rotation.

We therefore share Figure~\ref{fig:rotation-profiles} as a controlled sign-pattern comparison, rather than as a fit to any individual snapshot. The first row shows one fiducial member of the scale-free family: the type-II distribution functions defined by \citet{deBruijne+96}, which provide a convenient treatment of the velocity-anisotropy parameter, combined with a logarithmic potential,\footnote{This choice implies a flat circular-speed curve, as often observed in dark matter-dominated stellar systems.} isotropic velocities with $\beta=0$, a stellar-density slope of $2$,\footnote{This choice follows the formalism adopted in \cite{Vitral+24} to study dwarf spheroidal galaxies.} and intrinsic flattening $q=0.5$. The rotation profile is controlled by two parameters, $s$ and $t$, which set the fraction of rotating stars and the shape of the rotation profile, respectively. We choose $s=1$ and $t=1$, corresponding to maximal rotation with an intermediate profile shape. These choices define a simple, high-contrast example -- changing them modifies the amplitudes and detailed curve shapes, but not the qualitative sign symmetries discussed below. The second row repeats the same exercise for the Lynden-Bell--Plummer kinematic baseline adopted in this work, evaluated at the fiducial projected distance $R_{\rm POS}=r_{\phi}/2$ from the centre of the system. 
Finally, the bottom row depicts a representative snapshot of the $M_{\rm halo}=10^{9}~\msun$ simulation, evaluated at the same relative radius.

The point of Figure~\ref{fig:rotation-profiles} is therefore not the precise amplitude of any one curve, but the ordered angular structure expected from stationary axisymmetric rotation. Inclination changes the projected amplitudes, and in some cases can suppress a component, but it does not scramble the sign sequence across projected quadrants and radii. With the convention shown in the figure, the $v_{\rm POSr}$ component follows the quadrant sequence $(-,+,-,+)$, while $v_{\rm LOS}$ follows $(-,+,+,-)$.\footnote{Changing the sense of rotation, or equivalently reversing the sign of the angular momentum, simply flips these signs.} The $v_{\rm POSt}$ component is simpler still, since its sign does not alternate around the projected body. Thus, although the amplitudes of the projected moments may vary with radius, inclination, and model parameters, a stationary axisymmetric rotator is not expected to display sign changes that vary erratically with both projected radius and position angle.

This contrasts with several of the simulated snapshots.\footnote{A full time sequence of these kinematic maps is provided as online material.} For example, the $v_{\rm POSr}$ map at $t=3.6$~Gyr in Figure~\ref{fig:posr-project-v-maps} and the $v_{\rm POSt}$ map at $t=1.2$~Gyr in Figure~\ref{fig:post-project-v-maps} show sign changes that vary across both projected radius and azimuth, rather than following the ordered angular patterns illustrated in Figure~\ref{fig:rotation-profiles}.\footnote{The stronger visual departures in the plane-of-sky components may be partly geometrical. The encounter-driven velocity shifts need not be intrinsically weaker along the line of sight, but they can be projected on top of a stronger dipole-like first moment, making local reversals less apparent than in the plane-of-sky radial and tangential fields.} In addition, a persistent misalignment between the photometric and kinematic axes is not a natural expectation for a single stationary axisymmetric rotating configuration, and would usually point either to a more complex geometry or to a perturbative history.

\subsubsection{Triaxial expectations}
\label{sssec:triaxiality}

A natural possibility is that the stellar system is not well described as axisymmetric. For instance, more general triaxial configurations can produce apparent misalignments between photometric and kinematic axes. To explore this possibility, we turn to the formalism introduced in the seminal work of \cite*{Franx+91}, who showed that statistical arguments based on projected observables can be used to assess whether an observed photometric--kinematic misalignment can plausibly arise from projection alone, or instead, if it points towards an intrinsic misalignment.

In particular, purely geometrical considerations show that, if the intrinsic angular momentum is aligned with the short axis of a triaxial spheroid, the joint distribution of projected ellipticity $\epsilon$ and projected photometric--kinematic misalignment $\Psi \equiv |\Delta \xi|$, $p(\epsilon,\Psi)$, is fully specified by intrinsic shape parameters (see equation~\ref{eq:triaxial_intrinsic_eps}). Such aligned triaxial models generally predict an anti-correlation between $\epsilon$ and $\Psi$, which \citet{Franx+91} characterised through the statistic
$\langle \epsilon \Psi \rangle / (\langle \epsilon \rangle \langle \Psi \rangle) - 1$ (cf. their figure~4).
Hence, despite intrinsic shape parameters being inaccessible through direct observations, one can compare the observed distribution of a population of galaxies with the corresponding $p(\epsilon,\Psi)$ (or, e.g., with the expected behaviour of their correlation statistic), and therefore assess whether the observed projected misalignments are consistent with an intrinsically aligned triaxial population.

Motivated by this formalism, which we describe in detail in Appendix~\ref{sec:triaxial_projection}, we fit each snapshot of a given mass model with an intrinsic triaxial Plummer profile. We then use the intrinsic shape parameters inferred for each mass model and each snapshot to construct the corresponding aligned prediction for $p(\epsilon,\Psi)$ empirically, through sampling over a unit sphere. In Figure~\ref{fig:triaxiality-misalignment}, we overlay these expectations on the $\epsilon$--$\Psi$ distributions measured from our subhalo experiments. Black points and the associated grey density contours show the aligned triaxial prediction, while the orange points show the simulated measurements. 
For context, we also report the
$\langle \epsilon \Psi \rangle / (\langle \epsilon \rangle \langle \Psi \rangle) - 1$
statistic computed over the full set of snapshots in each mass model. In a second row, we show the corresponding distributions of the inferred
intrinsic shape parameters (cf. Equations~\ref{eq:triaxial_intrinsic_eps}
and~\ref{eq:triaxial_T}).

The comparison reveals clear differences between the three mass models. The two lowest-mass models show no significant correlation between $\epsilon$ and $\Psi$, whereas the $10^{10}~\msun$ model yields
$\langle \epsilon \Psi \rangle / (\langle \epsilon \rangle \langle \Psi \rangle) - 1=-0.21$, closer to the expectations from \citet{Franx+91}. Moreover, the full distribution of $(\epsilon,\Psi)$ pairs agrees remarkably well with the aligned prediction for the most massive halo, while the lower-mass simulations show a statistically significant excess of systems with simultaneously high $\epsilon$ and $\Psi$. 
The lower panels further show that, with increasing halo mass, the intrinsic stellar distribution progressively departs from near-sphericity towards more elongated shapes, spanning triaxial, oblate- and prolate-like configurations alike. Taken together, while the $M_{\rm halo}=10^{10}~\msun$ model is comparatively consistent with the aligned case, the lower-mass $M_{\rm halo}=3\times10^{8}$ and $10^{9}~\msun$ models, admittedly more realistic, are not. 
The fact that the higher-mass models also display stronger $(v/\sigma)$ measurements is qualitatively consistent with the results of \citet{Ene+18}, who found, for a sample of 90 early-type galaxies, that stronger rotators tend to show smaller photometric--kinematic misalignments than weaker rotators.

An important caveat is that the snapshots entering each mass model do not constitute a set of strictly independent galaxies, since successive snapshots retain some memory of their preceding dynamical evolution. 
To test whether the trends seen in Figure~\ref{fig:triaxiality-misalignment} are driven primarily by this temporal dependence, we performed a separate experiment in which the same snapshot and halo mass were analysed across independent initial realisations of the subhalo population. These tests recover qualitatively similar behaviour, although with larger uncertainties because of the smaller sample size. This suggests that the main trends identified from the full snapshot sequence are not simply a consequence of repeated sampling of a correlated evolutionary track, and are instead representative of the broader stochastic response of the system to subhalo perturbations.
Overall, the analysis in this subsection indicates that the measured misalignments, especially in the lower-mass models, are not merely projection effects in intrinsically aligned triaxial systems, but reflect intrinsic departures from aligned equilibrium expectations.

\subsubsection{Intrinsic streaming coherence}
\label{sssec:preferred-streaming-direction}

If neither an axisymmetric equilibrium configuration nor an intrinsically aligned triaxial model provides a satisfactory description of the geometry and kinematics of our subhalo experiments, at least for the lower-mass models, a more basic question remains: do these systems rotate at all, in the broader sense of possessing a preferred direction of stellar streaming, or are we seeing apparent effects caused by projection? We address this question directly in three dimensions, without imposing a particular equilibrium geometry.
Specifically, we ask whether the stellar angular momenta show a coherent
orientation and, if so, whether this intrinsic coherence is related to the
rotation diagnostics inferred from our projected Lynden-Bell--Plummer fits.

Appendix~\ref{sec:angular_momentum_coherence} introduces two complementary
statistics, $C$ and $R$. The former measures the coherence of the stellar
angular momenta while retaining their individual amplitudes, whereas the
latter considers only their directions and therefore gives equal weight to
each well-defined angular-momentum vector. Both satisfy
$0\leq (C,R) \leq1$: values close to zero correspond to little net directional
coherence, while values approaching unity indicate angular momenta strongly
concentrated around a common axis. In an approximately pressure-supported
system with largely disordered motions, both quantities are therefore
expected to remain small, although consistently positive values within uncertainties can still provide
evidence for coherent streaming.

Figure~\ref{fig:correlations-streaming} compares these three-dimensional
diagnostics with the quantities inferred from our Lynden-Bell--Plummer
formalism, using the intermediate-mass model as a representative example.
The upper panel shows $C$ as a function of $R$. Their clear positive
relation indicates that snapshots with stronger angular-momentum-weighted
coherence also show a stronger alignment of the individual angular-momentum
directions. The inferred streaming signal is therefore not driven solely by
a small number of stars carrying unusually large angular momenta, but is
shared more broadly by the stellar population. In this limited but useful
sense, the simulations do develop a preferred sense of rotation even though
their velocity fields might not be described by a stationary equilibrium configuration.

The middle panel then asks whether this intrinsic coherence is reflected in
our projected rotation measurement. The positive correlation between
$\log(v/\sigma)$ and $R$ shows that snapshots with a more strongly aligned
three-dimensional angular-momentum field also tend to be assigned larger
rotation amplitudes by the projected Lynden-Bell--Plummer fit. Thus,
although $(v/\sigma)$ compresses a spatially complex velocity field into a
global projected quantity, it nevertheless traces a genuine component of the
underlying coherent streaming.

Finally, the lower panel compares two measurements of projected
misalignment. We project both the intrinsic short axis inferred from the three-dimensional triaxial fit and the preferred angular-momentum axis onto the plane of the sky, and define $\Psi_{\rm proj}$ as the angle between these two projected directions.\footnote{For an intrinsically aligned rotator in the sense of \cite{Franx+91}, $\Psi_{\rm proj} \equiv \Psi$.} We then compare this quantity with the independently fitted photometric--kinematic
misalignment $|\Delta \xi|$.
Their strong correlation, with ${\rm d} \Psi_{\rm proj} / {\rm d} |\Delta \xi| \approx 1$, shows that the
misalignment recovered by the Lynden-Bell--Plummer model tracks the
orientation of the underlying three-dimensional streaming axis rather than
arising simply from an accidental projection of an otherwise incoherent
velocity field.

Taken together, these comparisons suggest a consistent picture: the systems
are not well described as stationary rotators about a single equilibrium
symmetry axis, but neither are their fitted rotation signals merely
projection artefacts. Instead, subhalo perturbations generate a genuine
coherent component of stellar angular momentum, superposed on the
spatially localised and time-dependent streaming structure discussed above.

\subsubsection{On the origin of measured asymmetries}
\label{sssec:asymmetry-origin}

Within our controlled experiments, departures from symmetry are generated by the imposed subhalo population. The time sequence of the kinematic maps shows that strong local streaming features are often associated with the passage of massive subhaloes. These encounters perturb the local velocity distribution in all three projected components, producing signatures that are localised in both space and time. 
The fitted $(v/\sigma)$ values therefore capture a genuine leading-order rotation-like component, but they compress the velocity field into a single global amplitude and do not describe its localised spatial structure.
The residual power-spectra shown in Figure~\ref{fig:ps-halo-mass} retain part of this additional structure by measuring the characteristic scales and amplitudes of the departures in the $\chi_u$ fields, even though these residuals are defined with respect to our adopted Lynden-Bell--Plummer kinematic baseline.

The dependence on halo mass then reflects the strength of the same underlying mechanism. What changes across the mass models is the amplitude of the fluctuations of the combined force induced by the subhalo population. More massive subhaloes transfer more energy during their passages and produce stronger local shifts in the velocity distribution, which explains why the $M_{\rm halo}=10^{10}~{\rm M}_{\odot}$ experiment develops the largest rotation-like amplitudes and the most pronounced departures, at least in a statistical sense. In the lower-mass cases, the same process still induces measurable streaming motions, but with smaller amplitudes relative to the velocity dispersion, which itself is shaped by the heating and expansion described in P25. In this sense, the simulated galaxies do rotate, but only in a general sense: they acquire non-zero rotation diagnostics, while their full velocity fields retain spatially localised signatures of the subhaloes' orbital histories.

\subsection{Alternative perturbations}
\label{ssec:alternative-perturbations}

The previous subsection argued that the streaming motions in our simulations are not fully captured by the usual picture of stationary rotation around a single smooth symmetry axis. A natural follow-up question is whether similar signatures could be produced by other perturbative mechanisms. A complete answer would require a dedicated comparison suite, including tides, mergers, and more general equilibrium models. Here, we instead use existing work to clarify which features are likely to be generic, and which may be more characteristic of the subhalo-driven experiments studied in this paper.

In dwarf spheroidal galaxies, the available data have so far mostly motivated fits for a global rotation amplitude and projected rotation axis, rather than detailed searches for spatially structured departures from such baselines. Related systems, however, already show that non-trivial streaming can arise through several channels. For instance, \citet*{Tiongco+18} studied rotating star clusters evolving in an external tidal field and found that tidal torques can induce precession and nutation of the rotation axis, radial twists in its orientation, and, in some cases, apparent counterrotation between inner and outer regions. This is an important warning: counterrotation or a radially varying kinematic axis is not unique to subhalo perturbations. At the same time, the tidal response in those models is organised by the external host field and by the orbital geometry of the cluster. 
This suggests a potentially useful distinction: perturbations driven by an external tidal field may retain some degree of large-scale symmetry, with kinematic twists organised by the host-orbit geometry and varying relatively smoothly with radius. Subhalo passages, by contrast, can produce more localised and directionally heterogeneous features, because the perturbing potential is tied to the instantaneous positions and orbital histories of individual subhaloes.
This distinction remains qualitative, especially because the clusters of \citet{Tiongco+18} are not embedded in an extended dark matter halo. Such a halo may reduce or redistribute the stellar response to an external tide \citep[e.g.][]{Vitral&Boldrini22}, although the strength of this effect should depend on the halo structure, orbit, and tidal history.

Merger and flyby scenarios provide another useful comparison. In the giant elliptical IC~1459, \citet{Franx&Illingworth98} found a rapidly counterrotating stellar core and discussed a merger or accretion event as a plausible origin. Similarly, \citet{DeRijcke+04} and \citet*{Geha+05} identified kinematically decoupled cores in dwarf ellipticals, and argued that external interactions, including flybys and minor mergers, can transfer enough angular momentum to generate such peculiar rotation profiles.
These cases show that strong changes in the rotation field, including counterrotation or rotation bumps along the radial profile, can be long-lived fossils of past interactions. They therefore reinforce the point that complex projected rotation is not, by itself, a unique diagnostic of present-day disequilibrium or of subhalo interactions. At the same time, mergers and accretion events may leave additional traces that are not expected from purely dark perturbers, such as chemically distinct stellar populations around such streaming features, or coherent and distinct structures in integrals-of-motion space \citep{Helmi&deZeeuw00}. When available, chemical abundances and phase-space clustering diagnostics could therefore help distinguish accreted stellar components from subhalo-induced perturbations.

Additionally, the subhalo experiments studied here do not merely generate a global twist or a central kinematically decoupled component. They produce streaming features that are often localised in both space and time, and whose morphology follows the changing orbital positions of massive subhaloes. This locality is visible in the velocity maps and is partly encoded in the residual power-spectra of the $\chi_u$ fields. The power-spectrum is therefore not only a detection tool for departures from the adopted Lynden-Bell--Plummer baseline: it may also provide a way to compare the characteristic spatial scales of different perturbative mechanisms.

These considerations suggest a practical route forward. Future simulations of dwarf spheroidals subjected to external tides, flybys, mergers, and subhalo populations should be analysed with the same projected diagnostics: the leading-order rotation signal $(v/\sigma)$, the photometric--kinematic misalignment $|\Delta \xi|$, the symmetry behaviour of the three mean velocity fields, and the power-spectra of the residual $\chi_u$ maps. Such a comparison would make it possible to determine whether the irregular, patchy plane-of-sky streaming found here is genuinely distinctive of subhalo interactions, or whether it can also arise from other perturbative histories.

\subsection{Observational prospects}
\label{ssec:observational-prospects}

Having discussed the nature of the streaming signatures in our simulations, and how they might be distinguished from other perturbative mechanisms, we finally ask whether the required observational data are available, or likely to become available in the near future. Current spectroscopic surveys of Milky Way dwarf spheroidal galaxies have already delivered large catalogues of line-of-sight velocities for many of the classical satellites \citep{Walker+23,Tolstoy+23,Geha+26}. These data are sufficient to measure global rotation amplitudes and projected kinematic axes in the most massive systems, as discussed in Section~\ref{ssec:delta-xi_v-over-sigma}. However, the more demanding diagnostics introduced in this work, especially the residual $\chi_u$ power-spectra, benefit strongly from larger tracer samples. As shown in Figure~\ref{fig:ps-stellar-count}, samples approaching $\mathcal{O}\left(10^{4}\right)$ well-measured velocities provide a much clearer route to detecting spatially structured kinematic departures from the adopted smooth rotating baseline.

Proper-motion measurements provide the complementary plane-of-sky information needed to turn this into a genuinely three-dimensional projected test. In this respect, the \gaia\ mission is central \citep{Gaia+16}. The current DR3 data already provide invaluable astrometric constraints, but their uncertainties and systematics remain too large to map localised internal proper-motion features in most dwarf spheroidals \citep{MartinezGarcia+21}. Gaia DR4 is expected to improve this situation by extending the astrometric time baseline and improving the quality of the proper-motion catalogue, potentially enabling robust internal plane-of-sky kinematic characterisations for the brightest and best-sampled systems. DR5 should extend these prospects further \citep{McKinnon&vanderMarel26}, bringing the possibility of combining line-of-sight velocities with statistically useful samples of proper motions for many of the classical dwarfs.

Although current data are not yet sufficient to probe all of the signatures identified in this work simultaneously, ongoing observational efforts are increasingly defining a clear and effective pathway towards that goal.
Continued spectroscopic campaigns can push line-of-sight catalogues towards the tracer counts required for robust residual power-spectrum measurements, while improved \gaia\ astrometry will open access to the complementary $v_{\rm POSr}$ and $v_{\rm POSt}$ dimensions, which are especially diagnostic in our simulations.\footnote{In this sense, pairing \gaia\ astrometry with deep, wide-field observations from \textit{Euclid} and the \textit{Nancy Grace Roman Space Telescope} will also be highly valuable \citep{Bedin25}.}
Together, these data will allow future studies to apply the full set of diagnostics developed here: $(v/\sigma)$, $|\Delta \xi|$, symmetry tests of the projected mean velocity fields, and the power-spectra of the residual $\chi_u$ maps. 
In doing so, it will be important to model observational perspective velocity gradients, which can mimic or blur intrinsic rotation signatures in extended dwarf galaxies \citep{vanderMarel+02,Walker+08,Kaplinghat&Strigari08}.
Complementary chemical information may further help separate perturbations caused by dark subhaloes from those associated with mergers or accreted stellar populations. In this way, the framework developed here connects increasingly large dwarf spheroidal kinematic data sets to population-level tests of dark substructure.

\section{Conclusion}
\label{sec:conclusion}

In this work, we have developed the velocity-space counterpart of the spatial analysis presented in Paper~I. Building on the controlled simulations of P25, in which repeated dark subhalo passages heat and expand an initially compact stellar tracer population, Paper~I showed that the projected stellar distribution remains well described at leading order by a smooth axisymmetric Plummer profile, while its residual density field contains coherent Fourier-space signatures of the subhalo population. Here, we asked whether analogous information is also present in the projected velocity field of the same systems.

To address this question, we fitted each simulation snapshot with a physically motivated smooth rotating Lynden-Bell--Plummer kinematic baseline, constructed normalised residual fields $\chi_u$ defined in equation~\ref{eq:methods_velocity_chi}, for the three projected velocity components, and analysed their power-spectra. The main conclusions are as follows.

\begin{itemize}
\item Subhalo perturbations induce transient streaming motions within the galaxy, which may appear as rotation-like signals despite being initialised without intrinsic rotation. In line-of-sight velocity alone, several snapshots resemble the classical blueshift--redshift pattern commonly associated with rotation.
\item The plane-of-sky velocity components (i.e. proper motions) reveal a more complex picture. The $v_{\rm POSr}$ maps display alternating regions of expansion and contraction, while the $v_{\rm POSt}$ maps show alternating clockwise and anticlockwise streaming regions. These patterns depart from the regular sign structures expected for stationary axisymmetric rotation, as illustrated by the scale-free and Lynden-Bell--Plummer reference models in Figure~\ref{fig:rotation-profiles}.
\item The residual power-spectra of the $\chi_u$ fields provide a compact quantitative description of these departures. The strength of the spectral excess increases with parent halo mass, reflecting the larger potential fluctuations induced by the correspondingly more massive subhaloes. 
Compared with the density-based analysis of Paper~I, the subhalo-induced signal appears especially evident in velocity space, even in the lowest-mass halo model. Kinematic diagnostics therefore provide a particularly sensitive complement to spatial residuals in searches for dark substructure.
Conversely, decreasing the number of stellar tracers progressively suppresses the detectability of the signal, with robust recovery requiring samples approaching $\mathcal{O}\left(10^{4}\right)$ stellar counts.
\item The recovered rotation ratio $(v/\sigma)$ and the projected photometric--kinematic misalignment $|\Delta \xi|$ populate the range inferred in recent studies of classical Milky Way dwarf spheroidals, especially for models with $M_{\rm halo}\lesssim 10^{9}~{\rm M}_{\odot}$ (see Figure~\ref{fig:delta-xi_v-sigma}).
\item Projection effects in intrinsically aligned triaxial systems do not readily explain the photometric--kinematic misalignments recovered in the two lower-mass models. These cases show an excess of snapshots with simultaneously high projected ellipticity and high misalignment relative to the aligned triaxial expectation, suggesting the development of a non-zero intrinsic misalignment. By contrast, the $M_{\rm halo}=10^{10}~{\rm M}_{\odot}$ model is more consistent with the aligned triaxial case.
\item The streaming motions generated by subhaloes should therefore not be interpreted simply as classical stationary rotation in equilibrium. They contain a genuine leading-order component of coherent stellar angular momentum, which we tested through intrinsic angular-momentum coherence diagnostics, but this component is embedded in spatially irregular velocity structures that retain information about the local orbital histories of massive subhaloes.
\item Other mechanisms, including tides, flybys, mergers, and more general equilibrium configurations, could also generate complex projected kinematics. Counterrotation, kinematic-axis twists, and rotation bumps are therefore not unique signatures of subhaloes. A promising distinction is that external tides may retain large-scale geometrical ordering, while subhalo passages can generate more localised and directionally heterogeneous perturbations.\footnote{Such differences in characteristic spatial scale are precisely the kind of information captured by the residual power-spectrum comparison developed in Section~\ref{ssec:power-spectrum-results}.}
Mergers or accretion events, on the other hand, may leave additional diagnostics, such as chemically distinct stellar populations or coherent structures in integrals-of-motion space, which are not necessarily expected from purely dark perturbers. These qualitative distinctions should be tested with dedicated comparison simulations and, where possible, with joint chemo-dynamical data.
\end{itemize}

The framework developed here opens several natural follow-up directions. First, the velocity-space diagnostics introduced in this paper should be applied to simulations including external tides, baryonic mergers, and more general equilibrium configurations, using the same residual-map, power-spectrum, projected-misalignment, and angular-momentum-coherence machinery. Secondly, the kinematic signatures studied here can be combined with the spatial signatures identified in Paper~I. In particular, cross-correlations between density residuals and the $\chi_u$ velocity-residual fields may provide a more discriminating probe of whether spatial and kinematic fluctuations trace the same underlying subhalo population. Finally, complementary information from chemistry and integrals-of-motion space may help separate dark perturbers from accreted stellar components.

From an observational perspective, current dwarf spheroidal data sets are not yet sufficient to probe all of these diagnostics simultaneously. Nevertheless, the pathway is becoming increasingly promising. Continued spectroscopic campaigns will enlarge line-of-sight velocity samples, while improved \gaia\ astrometry and complementary deep, wide-field observations should open access to the plane-of-sky velocity dimensions that are especially diagnostic in our simulations. As these data sets approach the tracer counts explored here, the methods developed in this paper offer a practical route from increasingly detailed phase-space maps to population-level tests of dark substructure and, more generally, to assessing whether dwarf spheroidals are close to dynamical equilibrium at present -- this broader question is important because inferences about their dark matter content usually rely on equilibrium-based dynamical modelling. In this sense, the internal kinematics of dwarf spheroidals becomes a powerful observable for constraining not only the abundance and orbital structure of dark subhaloes, but ultimately the small-scale nature of dark matter itself.

\section{Acknowledgments}

% We thank the anonymous referee for providing insightful comments that improved the clarity of the manuscript.
EV acknowledges funding from the Royal Society, under the Newton International Fellowship programme (NIF\textbackslash R1\textbackslash 241973).
MGW acknowledges support from the National Science Foundation (NSF) grant AST-2206046. %Support for program JWST-AR-02352.001-A was provided by NASA through a grant from the Space Telescope Science Institute, which is operated by the Association of Universities for Research in Astronomy, Inc., under NASA contract NAS 5-03127. This material is based upon work supported by the National Aeronautics and Space Administration under Grant/Agreement No. 80NSSC24K0084 as part of the Roman Large Wide Field Science program funded through ROSES call NNH22ZDA001N-ROMAN.

We are grateful to Shweta Dalal for insightful discussions on numerical methods for non-uniform Fourier transforms and data visualisation, as well as to Anna Lisa Varri for useful references and suggestions on rotation in stellar systems. We also thank Rafaelle Pascale and José María Arroyo-Polonio for sharing their estimates of $(v/\sigma)$ and $|\Delta \xi|$ from \citet{Pascale+26} and \citet{Arroyo-Polonio+24}, respectively. Rafaelle Pascale additionally provided helpful suggestions on the presentation of our figures.

% This work has made use of data from the European Space Agency (ESA) mission \gaia\ (\url{https://www.cosmos.esa.int/gaia}), processed by the \gaia\ Data Processing and Analysis Consortium (DPAC, \url{https://www.cosmos.esa.int/web/gaia/dpac/consortium}). Funding for the DPAC has been provided by national institutions, in particular the institutions participating in the \gaia\ Multilateral Agreement.

The following software and tools were used extensively during the writing of this manuscript: {\sc Python} \citep{VanRossum09}, 
{\sc BALRoGO} \citep{Vitral21},
{\sc scalefree} \citep{deBruijne+96,Vitral+24},
{\sc Scipy} \citep{Jones+01},
{\sc Numpy} \citep{vanderWalt11},
{\sc Matplotlib} \citep{Hunter07},
{\sc Visual Studio Code} \citep{microsoft2026vscode},
and {\sc ChatGPT} \citep{openai2026chatgpt}.
%%%%%%%%%%%%%%%%%%%%%%%%%%%%%%%%%%%%%%%%%%%%%%%%%%
\section{Data Availability}

The data that support the plots within this paper and other findings of this study are available from the corresponding author upon request.

%%%%%%%%%%%%%%%%%%%% REFERENCES %%%%%%%%%%%%%%%%%%

% The best way to enter references is to use BibTeX:

\bibliographystyle{mnras}
\bibliography{src} % if your bibtex file is called example.bib

% Alternatively you could enter them by hand, like this:
% This method is tedious and prone to error if you have lots of references
%\begin{thebibliography}{99}
%\bibitem[\protect\citeauthoryear{Author}{2012}]{Author2012}
%Author A.~N., 2013, Journal of Improbable Astronomy, 1, 1
%\bibitem[\protect\citeauthoryear{Others}{2013}]{Others2013}
%Others S., 2012, Journal of Interesting Stuff, 17, 198
%\end{thebibliography}

%%%%%%%%%%%%%%%%%%%%%%%%%%%%%%%%%%%%%%%%%%%%%%%%%%

%%%%%%%%%%%%%%%%% APPENDICES %%%%%%%%%%%%%%%%%%%%%

\appendix

\section{Handling of rotation}
\label{sec:rotation_models}

\subsection{Mathematical formalism}
\label{ssec:mathematical_formalism}

\begin{figure}
    \centering
    \includegraphics[width=\hsize]{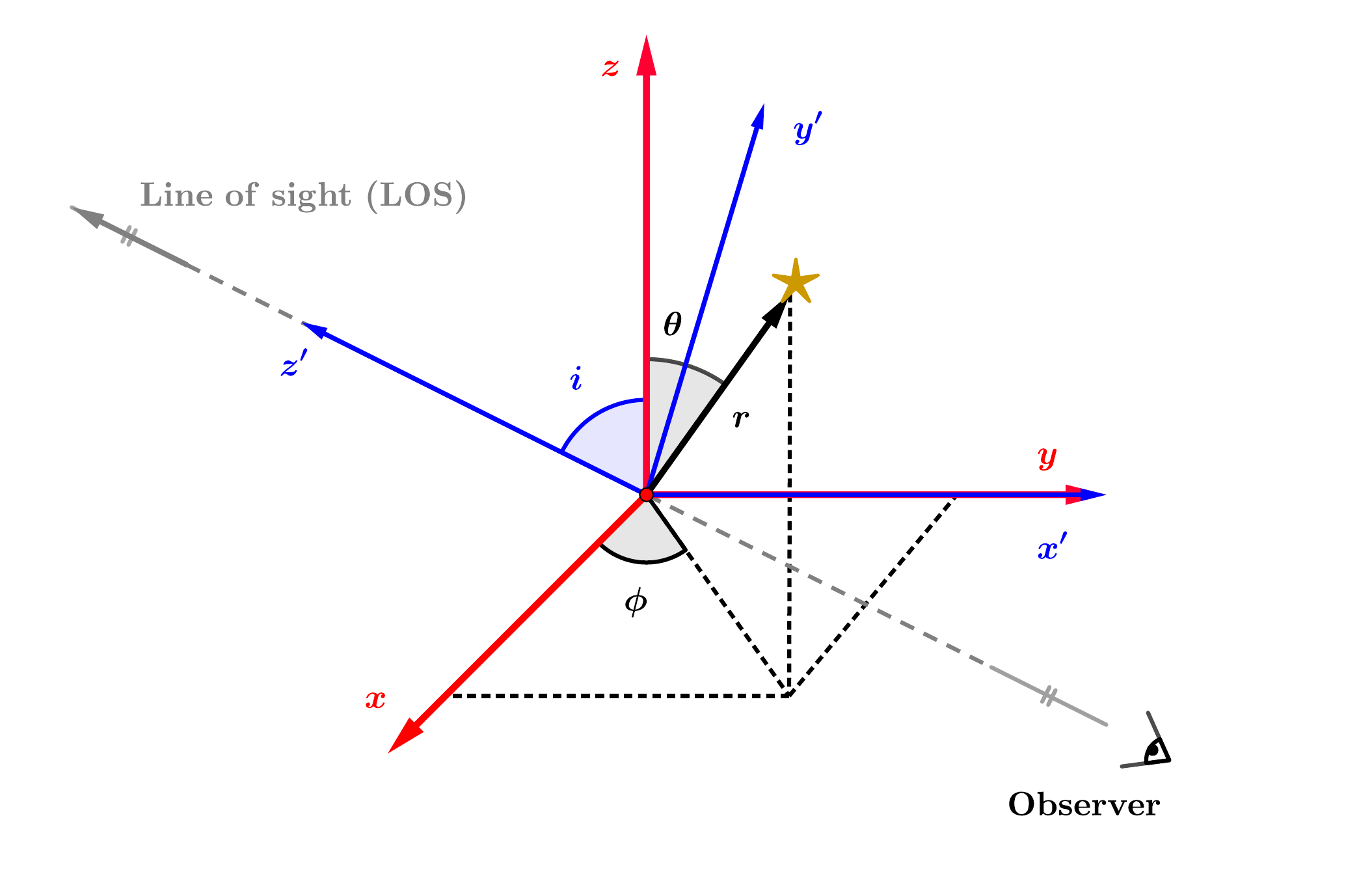}
    \caption{\textit{Geometry of the system.}
    Definition of the intrinsic coordinates \((x,y,z)\) and the projected
    coordinates \((x',y',z')\). The source is observed at an inclination
    angle \(i\), and the \(z'\)-axis is aligned with the line of sight.
    We adopt the same convention as presented in a similar figure by
    \protect\citet{Vitral+24}.}
    \label{fig:geom-frame}
\end{figure}

Here, we describe the mathematical formalism used to model ordered
rotation throughout this work. We follow the projection conventions of
\citet{Evans&deZeeuw94}, where \((x',y')\) are Cartesian coordinates in
the plane of the sky, aligned with the projected major and minor axes,
respectively, and \(z'\) is the line-of-sight coordinate. The
corresponding intrinsic coordinates are denoted by \((x,y,z)\), with
\(z\) aligned with the symmetry axis of the galaxy. These definitions
are illustrated in Figure~\ref{fig:geom-frame}.

We assume that the ordered motion is purely azimuthal about the
intrinsic symmetry axis and parametrise the mean streaming velocity as
\begin{equation}
\langle v_{\phi}\rangle
=
\frac{\omega R}{1+(r/r_{\phi})^{2}},
\label{eq:rotation_law}
\end{equation}
where $R=(x^{2}+y^{2})^{1/2}$ and
$r=(x^{2}+y^{2}+z^{2})^{1/2}$ are the intrinsic cylindrical and
spherical radii, respectively. The parameter $r_{\phi}$ sets the
radial scale at which the rotation curve departs from solid-body
behaviour, while $\omega$ determines the central angular-velocity
scale. Since $R\leq r$, the maximum streaming speed is reached in the
equatorial plane, where $R=r$, giving
$v_{\rm max}=\omega r_{\phi}/2$. In the limit
$r_{\phi}\rightarrow\infty$, equation~\ref{eq:rotation_law}
reduces to solid-body rotation, $\langle v_{\phi}\rangle=\omega R$.

This parametrisation is motivated by the rotating collisionless models
of \citet[][see their equation~56]{LyndenBell67}, developed in the
context of elliptical galaxies, in which incomplete violent relaxation
produces approximately solid-body streaming in the central regions and
declining mean azimuthal velocities at larger radii. Although not
originally motivated by the dynamics of dwarf spheroidal galaxies,
these galaxies provide plausible analogues of pressure-supported,
collisionless systems that can retain coherent angular momentum.
\footnote{Here, relaxation refers broadly to collisionless phase mixing and orbital diffusion, rather than to two-body relaxation. Classical dwarf spheroidal galaxies are old, dark matter-dominated systems whose stellar components have evolved over many internal dynamical times and may have experienced time-dependent gravitational potentials during halo assembly \citep{White96}, baryonic feedback \citep{Pontzen&Governato12}, tidal interactions \citep{Mayer+01}, and encounters with dark subhaloes \citep{Penarrubia+25}. 
In this context, their stellar populations may therefore evolve towards a coarse-grained collisionless equilibrium of the type discussed by \citeauthor{LyndenBell67} (\citeyear{LyndenBell67}, which we use here as an equilibrium-inspired null model for smooth rotation), while retaining ordered rotation as a relic of their dynamical history.}
Equation~\ref{eq:rotation_law} therefore provides a simple,
equilibrium-inspired description of coherent rotation and a natural
smooth baseline for the models considered in this work, against which
departures induced by subhalo-driven perturbations may be quantified.

For an axisymmetric system, the mean motions in the intrinsic radial
and vertical directions vanish,
\begin{equation}
    \langle v_R\rangle=\langle v_z\rangle=0.
\end{equation}
We define the density-weighted projection of a quantity \(X\) along the
line of sight as
\begin{equation}
    \langle X\rangle_{\rm p}
    \triangleq
    \frac{1}{\Sigma}
    \int_{-\infty}^{+\infty}
    {\rm d}z'\,\rho X,
    \label{eq:projected_mean_definition}
\end{equation}
where \(\Sigma\) is the projected surface density. Throughout this
work, positive line-of-sight motion is defined along the positive
\(z'\)-direction, such that
\begin{equation}
    \langle v_{\rm LOS}\rangle
    \triangleq
    \langle v_{z'}\rangle_{\rm p}.
    \label{eq:zprime-los-definition}
\end{equation}
The sign of \(\langle v_{\rm LOS}\rangle\) therefore differs from the velocity convention adopted by
\citet{Evans&deZeeuw94}.

\subsection{Geometry, density, and finite-interval integrals}
\label{ssec:rotation_geometry_integrals}

Following equation~A1 of \citet{Evans&deZeeuw94}, the intrinsic and
projected coordinates are related by
\begin{equation}
    x=-y'\cos i+z'\sin i,
    \qquad
    y=x',
    \qquad
    z=y'\sin i+z'\cos i.
    \label{eq:coordinate_transform_rotation}
\end{equation}
The intrinsic spherical radius is therefore
\begin{equation}
    r^{2}
    =
    (x')^{2}+(y')^{2}+(z')^{2},
    \label{eq:spherical_radius_projected}
\end{equation}
while the cylindrical radius entering the rotation law is
\begin{equation}
    R^{2}
    =
    x^{2}+y^{2}
    =
    \left(-y'\cos i+z'\sin i\right)^{2}
    +(x')^{2}.
    \label{eq:cylindrical_radius_projected}
\end{equation}

We assume that the intrinsic stellar density follows the deprojection
of an axisymmetric Plummer profile,
\begin{equation}
    \rho(x,y,z)
    =
    \frac{3N_{\infty}}{4\pi r_{\rm scale}^{3}q}
    \left[
    1+
    \frac{x^{2}+y^{2}+z^{2}/q^{2}}{r_{\rm scale}^{2}}
    \right]^{-5/2},
    \label{eq:intrinsic_plummer_density}
\end{equation}
where \(N_{\infty}\) is the total number of tracers,
\(r_{\rm scale}\) is the Plummer scale radius, and \(q\) is the intrinsic axial ratio. 
This density should be understood as an analytic modelling choice, rather than as a fully self-consistent consequence of the Lynden-Bell distribution function adopted for the velocity moments. In a strictly self-consistent construction, the stellar density would follow from the distribution function and the assumed potential through $\rho_\star=\int d^3 v f_\star(r,v)$. Here, we instead retain the Plummer form because Paper~I showed that it describes the projected stellar distribution well at leading order, and because it keeps the projection integrals analytically tractable. The resulting Lynden-Bell--Plummer model should therefore be interpreted as a physically motivated smooth rotating kinematic baseline, but not as a self-consistent equilibrium model.

The projected axial ratio
\(q_{\rm p}=1-\epsilon\) is related to \(q\) and \(i\) through
\begin{equation}
    q_{\rm p}^{2}
    =
    \cos^{2}i+q^{2}\sin^{2}i,
    \label{eq:projected_flattening}
\end{equation}
such that the corresponding projected surface density becomes
\begin{equation}
    \Sigma(x',y')
    =
    \frac{N_{\infty}}{\pi r_{\rm scale}^{2}q_{\rm p}}
    \left[
    1+
    \frac{1}{r_{\rm scale}^{2}}
    \left(
    (x')^{2}
    +
    \frac{(y')^{2}}{q_{\rm p}^{2}}
    \right)
    \right]^{-2}.
    \label{eq:projected_surface_density}
\end{equation}
For later convenience, we define
\begin{equation}
    \mathcal{S}(x',y')
    \triangleq
    1+
    \frac{1}{r_{\rm scale}^{2}}
    \left[
    (x')^{2}
    +
    \frac{(y')^{2}}{q_{\rm p}^{2}}
    \right].
    \label{eq:surface_density_shape_factor}
\end{equation}

The ellipsoidal radius entering equation~\ref{eq:intrinsic_plummer_density}
can be written as
\begin{equation}
\begin{split}
    m^{2}
    &\triangleq
    x^{2}+y^{2}+\frac{z^{2}}{q^{2}}
    \\
    &=
    (x')^{2}
    +(y')^{2}
    \left(
    \cos^{2}i+\frac{\sin^{2}i}{q^{2}}
    \right)
    +(z')^{2}
    \left(
    \sin^{2}i+\frac{\cos^{2}i}{q^{2}}
    \right)
    \\
    &\quad
    +2y'z'\sin i\cos i
    \left(
    \frac{1}{q^{2}}-1
    \right).
\end{split}
\label{eq:ellipsoidal_radius_projected}
\end{equation}
We now define
\begin{align}
    a
    &\triangleq
    r_{\phi}^{2}+(x')^{2}+(y')^{2},
    \label{eq:a_def}
    \\
    b
    &\triangleq
    (x')^{2}
    +(y')^{2}
    \left(
    \cos^{2}i+\frac{\sin^{2}i}{q^{2}}
    \right),
    \label{eq:b_def}
    \\
    c
    &\triangleq
    \sin^{2}i+\frac{\cos^{2}i}{q^{2}},
    \label{eq:c_def}
    \\
    d
    &\triangleq
    2y'\sin i\cos i
    \left(
    \frac{1}{q^{2}}-1
    \right).
    \label{eq:d_def}
\end{align}
Thus
\begin{equation}
    r^{2}+r_{\phi}^{2}=a+(z')^{2},
    \qquad
    m^{2}=b+dz'+c(z')^{2}.
    \label{eq:r_m_abcd}
\end{equation}
Completing the square in the density term, we further define
\begin{equation}
    z_{0}
    \triangleq
    -\frac{d}{2c},
    \qquad
    \lambda^{2}
    \triangleq
    \frac{4c(r_{\rm scale}^{2}+b)-d^{2}}{4c^{2}},
    \label{eq:z0_lambda_def}
\end{equation}
and take \(\lambda\) to be the positive square root. Along a fixed line
of sight, the density can then be written as
\begin{equation}
    \rho(z')
    =
    \frac{3N_{\infty}r_{\rm scale}^{2}}
    {4\pi q c^{5/2}}
    \frac{1}
    {
    \left[
    (z'-z_{0})^{2}+\lambda^{2}
    \right]^{5/2}
    }.
    \label{eq:rho_completed_square}
\end{equation}

All line-of-sight integrals required below can be expressed using the
same finite-interval transformation. We set
\begin{equation}
    z'
    =
    z_{0}+\lambda\tan\theta,
    \qquad
    t=\tan\frac{\theta}{2},
    \label{eq:z_theta_t_substitution}
\end{equation}
so that \(z'\in(-\infty,+\infty)\) maps to \(t\in(-1,+1)\). Equivalently,
\begin{equation}
    z'
    =
    \frac{
    z_{0}(1-t^{2})+2\lambda t
    }
    {1-t^{2}}.
    \label{eq:zprime_t_form}
\end{equation}
For compactness, we define
\begin{equation}
    U(t)
    \triangleq
    z_{0}(1-t^{2})+2\lambda t,
    \qquad
    V(t)
    \triangleq
    1-t^{2},
    \qquad
    W(t)
    \triangleq
    1+t^{2},
    \label{eq:UVW_def}
\end{equation}
and
\begin{equation}
    A(t)
    \triangleq
    aV^{2}(t)+U^{2}(t).
    \label{eq:A_t_def}
\end{equation}
The dimensionless integral family used throughout the projected first
and second moments is
\begin{equation}
    \mathcal{T}_{N}^{(M)}
    \triangleq
    \int_{-1}^{+1}
    \frac{
    2U^{N}(t)V^{2M+3-N}(t)
    }
    {
    W^{4}(t)A^{M}(t)
    }
    \,{\rm d}t.
    \label{eq:TNM_def}
\end{equation}
No symmetry assumption has been made, so odd-\(N\) integrals are
retained. The first moments depend on
\(\mathcal{T}_{0}^{(1)}\) and \(\mathcal{T}_{1}^{(1)}\), while the
second moments require
\begin{equation}
    \left\{
    \mathcal{T}_{0}^{(1)},
    \mathcal{T}_{1}^{(1)},
    \mathcal{T}_{2}^{(1)},
    \mathcal{T}_{0}^{(2)},
    \mathcal{T}_{1}^{(2)},
    \mathcal{T}_{2}^{(2)}
    \right\}.
    \label{eq:TNM_required}
\end{equation}

\subsection{Projected first moments}
\label{ssec:projected_first_moments}

Following equation~A4 of \citet{Evans&deZeeuw94}, and using our
line-of-sight sign convention, the projected first moments are
\begin{align}
    \langle v_{x'}\rangle_{\rm p}
    &=
    \frac{1}{\Sigma}
    \int_{-\infty}^{+\infty}
    {\rm d}z'\,
    \frac{x}{R}\,
    \rho\,
    \langle v_{\phi}\rangle,
    \label{eq:vxprime_projection_general}
    \\
    \langle v_{y'}\rangle_{\rm p}
    &=
    \frac{\cos i}{\Sigma}
    \int_{-\infty}^{+\infty}
    {\rm d}z'\,
    \frac{y}{R}\,
    \rho\,
    \langle v_{\phi}\rangle,
    \label{eq:vyprime_projection_general}
    \\
    \langle v_{z'}\rangle_{\rm p}
    &=
    -\frac{\sin i}{\Sigma}
    \int_{-\infty}^{+\infty}
    {\rm d}z'\,
    \frac{y}{R}\,
    \rho\,
    \langle v_{\phi}\rangle.
    \label{eq:vzprime_projection_general}
\end{align}
Since \(y=x'\), and since \(R\) cancels after substituting
equation~\ref{eq:rotation_law}, these expressions depend only on
\(\mathcal{T}_{0}^{(1)}\) and \(\mathcal{T}_{1}^{(1)}\).

It is useful to define the common first-moment prefactor
\begin{equation}
    \mathcal{A}(x',y')
    \triangleq
    \frac{
    3\omega r_{\phi}^{2}
    r_{\rm scale}^{4}q_{\rm p}
    }
    {
    4q\,c^{5/2}\lambda^{4}
    }
    \mathcal{S}^{2}(x',y').
    \label{eq:rotation_common_factor}
\end{equation}
Here, the factor \(N_{\infty}\) has cancelled after substituting
equation~\ref{eq:projected_surface_density}. The projected ordered
velocity field is then
\begin{align}
    \langle v_{z'}\rangle_{\rm p}
    &=
    -\mathcal{A}\,x'\sin i\,\mathcal{T}_{0}^{(1)},
    \label{eq:vzprime_final_compact}
    \\
    \langle v_{y'}\rangle_{\rm p}
    &=
    \mathcal{A}\,x'\cos i\,\mathcal{T}_{0}^{(1)},
    \label{eq:vyprime_final_compact}
    \\
    \langle v_{x'}\rangle_{\rm p}
    &=
    \mathcal{A}
    \left[
    -y'\cos i\,\mathcal{T}_{0}^{(1)}
    +
    \sin i\,\mathcal{T}_{1}^{(1)}
    \right].
    \label{eq:vxprime_final_compact}
\end{align}
Equation~\ref{eq:vyprime_final_compact} also gives
\begin{equation}
    \langle v_{y'}\rangle_{\rm p}
    =
    -\frac{\langle v_{z'}\rangle_{\rm p}}{\tan i}.
    \label{eq:vyprime_from_vzprime}
\end{equation}
For \(x'\neq0\), the first term in
equation~\ref{eq:vxprime_final_compact} may be expressed through
\(\langle v_{z'}\rangle_{\rm p}\), but the form above remains well
defined on the projected minor axis.

\subsection{Projected second-order moments}
\label{ssec:projected_second_moments}

We now extend the projected formalism to raw second-order velocity moments. The distribution function of the rotating model in \citet[][equation~56]{LyndenBell67} is locally Gaussian in the spherical velocity components and contains no cross terms. This allows one to identify the intrinsic spherical dispersions as
\begin{align}
\sigma_r = \frac{1}{\sqrt{\mu}}, \quad
\sigma_{\theta} &\equiv \sigma_{\phi}
=
\frac{r_{\phi}}{\sqrt{\mu\left(r_{\phi}^{2}+r^{2}\right)}}.
\label{eq:sigma-definition}
\end{align}
Here, $\mu$ is the Lagrange multiplier associated with conservation of total energy in the formalism of \citet{LyndenBell67}.\footnote{In the original notation of \citet{LyndenBell67}, this parameter is denoted by $\beta$. Here, we use $\mu$ instead, to avoid confusion with the velocity anisotropy parameter, which conventionally uses the same notation.} Physically, it can be interpreted as an inverse kinetic temperature per unit mass, or equivalently as an inverse velocity-dispersion scale. Together with the maximum streaming speed introduced in Section~\ref{ssec:mathematical_formalism}, this introduces the dimensionless rotation scale
\begin{equation}
\left(\frac{v}{\sigma}\right)
\triangleq
\left|\frac{v_{\rm max}}{\sigma_r}\right|
=
\left|\frac{\omega r_{\phi}\sqrt{\mu}}{2}\right|,
\label{eq:lb_v_over_sigma}
\end{equation}
which measures the relative importance of ordered streaming motion with respect to the constant radial dispersion scale, $\sigma_r$.

We also note that, within this formalism, the stellar velocity anisotropy follows the classical Osipkov--Merritt parametrisation \citep{Osipkov79,Merritt85_df},
\begin{equation}
\beta(r)
\triangleq
1 - \frac{\sigma_{\theta}^{2}+\sigma_{\phi}^{2}}{2\sigma_{r}^{2}}
=
\frac{r^{2}}{r^{2}+r_{\phi}^{2}}.
\label{eq:anisotropy}
\end{equation}
The system is therefore isotropic at its centre and becomes increasingly radially anisotropic at large radii, approaching purely radial orbits asymptotically. Interestingly, Figure~6 of \citet{Penarrubia+25} shows that our dwarf galaxy realisations exhibit a similar radial trend throughout their evolution, thus adding verisimilitude to our adopted rotation model. 

After rotating the diagonal
second-moment tensor from spherical to cylindrical coordinates, and
using \(\langle v_R\rangle=\langle v_z\rangle=0\), we obtain
\begin{align}
    \langle v_R^2\rangle
    &=
    \frac{R^{2}+r_{\phi}^{2}}
    {\mu(r^{2}+r_{\phi}^{2})},
    \label{eq:MR_lb}
    \\
    \langle v_z^2\rangle
    &=
    \frac{z^{2}+r_{\phi}^{2}}
    {\mu(r^{2}+r_{\phi}^{2})},
    \label{eq:Mz_lb}
    \\
    \langle v_Rv_z\rangle
    &=
    \frac{Rz}
    {\mu(r^{2}+r_{\phi}^{2})},
    \label{eq:MRz_lb}
    \\
    \langle v_{\phi}^{2}\rangle
    &\equiv
    \sigma_{\phi}^{2}
    +
    \langle v_{\phi}\rangle^{2}
    =
    \frac{r_{\phi}^{2}}
    {\mu(r^{2}+r_{\phi}^{2})}
    +
    \left(
    \frac{\omega Rr_{\phi}^{2}}
    {r^{2}+r_{\phi}^{2}}
    \right)^{2},
    \label{eq:Mphi_lb}
\end{align}
The first three quantities are both dispersions and raw second moments,
whereas \(\langle v_{\phi}^{2}\rangle\) contains both the azimuthal
dispersion and the contribution from ordered streaming. 

The local projected second moments follow from the projected velocity
transformation of \citet[][equation~A5]{Evans&deZeeuw94}. We then
integrate them along the line of sight, as in their equation~A7. After
substituting equations~\ref{eq:MR_lb}--\ref{eq:Mphi_lb} and
collecting powers of \(z'\), all projected raw second moments can be
written in terms of the same \(\mathcal{T}_{N}^{(M)}\) integrals defined
above in equation~\ref{eq:TNM_def}. 
% To avoid introducing another normalisation, we use the common
% factor
% \begin{equation}
%     \frac{\mathcal{A}}{\omega r_{\phi}^{2}}
%     =
%     \frac{
%     3r_{\rm scale}^{4}q_{\rm p}
%     }
%     {
%     4q\,c^{5/2}\lambda^{4}
%     }
%     \mathcal{S}^{2}(x',y'),
%     \label{eq:second_moment_common_factor_from_A}
% \end{equation}
% where the equality on the right-hand side gives the explicit expression
% and should be used when considering the formal \(\omega\rightarrow0\)
% limit.

In this context, the diagonal raw second moments are
\begin{align}
    \langle v_{x'}^{2}\rangle_{\rm p}
    &=
    \frac{\mathcal{A}}{\omega r_{\phi}^{2}}
    \Bigg\{
    \frac{(x')^{2}+r_{\phi}^{2}}{\mu}
    \mathcal{T}_{0}^{(1)}
    \nonumber\\
    &\quad
    +
    \omega^{2}r_{\phi}^{4}
    \Big[
    (y')^{2}\cos^{2}i\,\mathcal{T}_{0}^{(2)}
    \nonumber\\
    &\quad\quad
    -
    2y'\sin i\cos i\,\mathcal{T}_{1}^{(2)}
    +
    \sin^{2}i\,\mathcal{T}_{2}^{(2)}
    \Big]
    \Bigg\},
    \label{eq:vxprime2_projected_TNM}
    \\
    \langle v_{y'}^{2}\rangle_{\rm p}
    &=
    \frac{\mathcal{A}}{\omega r_{\phi}^{2}}
    \left[
    \frac{(y')^{2}+r_{\phi}^{2}}{\mu}
    \mathcal{T}_{0}^{(1)}
    +
    (x')^{2}\omega^{2}r_{\phi}^{4}\cos^{2}i\,
    \mathcal{T}_{0}^{(2)}
    \right],
    \label{eq:vyprime2_projected_TNM}
    \\
    \langle v_{z'}^{2}\rangle_{\rm p}
    &=
    \frac{\mathcal{A}}{\omega r_{\phi}^{2}}
    \left[
    \frac{
    \mathcal{T}_{2}^{(1)}
    +
    r_{\phi}^{2}\mathcal{T}_{0}^{(1)}
    }{\mu}
    +
    (x')^{2}\omega^{2}r_{\phi}^{4}\sin^{2}i\,
    \mathcal{T}_{0}^{(2)}
    \right],
    \label{eq:vzprime2_projected_TNM}
\end{align}
while the mixed raw second moments are
\begin{align}
    \langle v_{x'}v_{y'}\rangle_{\rm p}
    &=
    \frac{\mathcal{A}}{\omega r_{\phi}^{2}}
    \Bigg\{
    \frac{x'y'}{\mu}
    \mathcal{T}_{0}^{(1)}
    \nonumber\\
    &\quad
    +
    x'\omega^{2}r_{\phi}^{4}
    \left[
    -y'\cos^{2}i\,\mathcal{T}_{0}^{(2)}
    +
    \sin i\cos i\,\mathcal{T}_{1}^{(2)}
    \right]
    \Bigg\},
    \label{eq:vxprime_vyprime_projected_TNM}
    \\
    \langle v_{x'}v_{z'}\rangle_{\rm p}
    &=
    \frac{\mathcal{A}}{\omega r_{\phi}^{2}}
    \Bigg\{
    \frac{x'}{\mu}
    \mathcal{T}_{1}^{(1)}
    \nonumber\\
    &\quad
    +
    x'\omega^{2}r_{\phi}^{4}
    \left[
    y'\sin i\cos i\,\mathcal{T}_{0}^{(2)}
    -
    \sin^{2}i\,\mathcal{T}_{1}^{(2)}
    \right]
    \Bigg\},
    \label{eq:vxprime_vzprime_projected_TNM}
    \\
    \langle v_{y'}v_{z'}\rangle_{\rm p}
    &=
    \frac{\mathcal{A}}{\omega r_{\phi}^{2}}
    \left[
    \frac{y'}{\mu}
    \mathcal{T}_{1}^{(1)}
    -
    (x')^{2}\omega^{2}r_{\phi}^{4}\sin i\cos i\,
    \mathcal{T}_{0}^{(2)}
    \right].
    \label{eq:vyprime_vzprime_projected_TNM}
\end{align}
These are projected raw second moments. The corresponding projected
dispersions and covariances are obtained by subtracting the products of
the projected first moments:
\begin{equation}
    \sigma_{u,{\rm p}}^{2}
    \equiv
    \langle v_{u}^{2}\rangle_{\rm p}
    -
    \langle v_{u}\rangle_{\rm p}^{2},
    \qquad
    u\in\{x',y',z'\},
    \label{eq:projected_dispersion_from_raw}
\end{equation}
and
\begin{equation}
    {\rm Cov}(v_u,v_v)
    \equiv
    \langle v_uv_v\rangle_{\rm p}
    -
    \langle v_u\rangle_{\rm p}
    \langle v_v\rangle_{\rm p}.
    \label{eq:projected_covariance_from_raw}
\end{equation}
Thus, the complete projected first- and second-order velocity structure
is determined by the same finite-interval integral family
\(\mathcal{T}_{N}^{(M)}\), evaluated over \(t\in[-1,+1]\).

\subsection{Projected radial and tangential velocity moments}
\label{ssec:projected_posrt_moments}

The projected Cartesian velocity moments derived above can also be recast in a local polar basis on the plane of the sky, providing a more natural interpretation in contexts where radial and tangential kinematic components are of interest.
At a fixed projected position
$(x',y')$, we define the projected radius with respect to the system's spatial centre, and respective normalised scales,
\begin{equation}
R_{\rm POS}
\triangleq
\sqrt{(x')^{2}+(y')^{2}},
\qquad
\eta_{x'}
\triangleq
\frac{x'}{R_{\rm POS}},
\qquad
\eta_{y'}
\triangleq
\frac{y'}{R_{\rm POS}}.
\label{eq:pos_eta_def}
\end{equation}
The projected radial and tangential velocity components are then
\begin{subequations}
\begin{align}
v_{\rm POSr}
&=
\eta_{x'} v_{x'}+\eta_{y'} v_{y'}
=
\frac{x'v_{x'}+y'v_{y'}}{R_{\rm POS}},
\\
v_{\rm POSt}
&=
\eta_{y'} v_{x'}-\eta_{x'} v_{y'}
=
\frac{y'v_{x'}-x'v_{y'}}{R_{\rm POS}}.
\end{align}
\label{eq:posrt_velocity_def}
\end{subequations}
% Equivalently,
% \begin{equation}
% \begin{pmatrix}
% v_{\rm POSr}\\
% v_{\rm POSt}
% \end{pmatrix}
% =
% \mathbf{P}_{\rm POS}
% \begin{pmatrix}
% v_{x'}\\
% v_{y'}
% \end{pmatrix},
% \qquad
% \mathbf{P}_{\rm POS}
% =
% \begin{pmatrix}
% \eta_{x'} & \eta_{y'}\\
% \eta_{y'} & -\eta_{x'}
% \end{pmatrix}.
% \label{eq:posrt_matrix_def}
% \end{equation}
% This transformation is orthogonal, i.e.,
% $\mathbf{P}_{\rm POS}\mathbf{P}_{\rm POS}^{\rm T}=\mathbf{I}$, but
% it is position-dependent. It is therefore undefined exactly at
% $R_{\rm POS}=0$, where the projected radial and tangential directions are not uniquely defined.

Since $x'$ and $y'$ are fixed along a given line of sight, the
projected first moments transform linearly:
\begin{subequations}
\begin{align}
\langle v_{\rm POSr}\rangle_{\rm p}
&=
\eta_{x'}\langle v_{x'}\rangle_{\rm p}
+
\eta_{y'}\langle v_{y'}\rangle_{\rm p},
\\
\langle v_{\rm POSt}\rangle_{\rm p}
&=
\eta_{y'}\langle v_{x'}\rangle_{\rm p}
-
\eta_{x'}\langle v_{y'}\rangle_{\rm p}.
\end{align}
\label{eq:posrt_mean_general}
\end{subequations}
% Substituting equations~\ref{eq:vxprime_final_compact} and
% \ref{eq:vyprime_final_compact}, these become
% \begin{subequations}
% \begin{align}
% \langle v_{\rm POSr}\rangle_{\rm p}
% &=
% \frac{\mathcal{A}x'\sin i}{R_{\rm POS}}\,
% \mathcal{T}_{1}^{(1)},
% \label{eq:vposr_mean_compact}
% \\
% \langle v_{\rm POSt}\rangle_{\rm p}
% &=
% \frac{\mathcal{A}}{R_{\rm POS}}
% \left[
% y'\sin i\,\mathcal{T}_{1}^{(1)}
% -
% R_{\rm POS}^{2}\cos i\,\mathcal{T}_{0}^{(1)}
% \right].
% \label{eq:vpost_mean_compact}
% \end{align}
% \label{eq:posrt_mean_compact}
% \end{subequations}

For the second-order moments, 
% define the projected Cartesian raw moments
% \begin{equation}
% M_{xx}\triangleq\langle v_{x'}^{2}\rangle_{\rm p},
% \qquad
% M_{yy}\triangleq\langle v_{y'}^{2}\rangle_{\rm p},
% \qquad
% M_{xy}\triangleq\langle v_{x'}v_{y'}\rangle_{\rm p}.
% \label{eq:cartesian_raw_moments_xy}
% \end{equation}
the raw projections in the $({\rm POSr}, {\rm POSt})$ basis
are
\begin{subequations}
\begin{align}
\langle v_{\rm POSr}^{2}\rangle_{\rm p}
&=
\eta_{x'}^{2} \langle v_{x'}^{2}\rangle_{\rm p}
+
\eta_{y'}^{2} \langle v_{y'}^{2}\rangle_{\rm p}
+
2\eta_{x'}\eta_{y'} \langle v_{x'}v_{y'}\rangle_{\rm p},
\label{eq:vposr2_raw}
\\
\langle v_{\rm POSt}^{2}\rangle_{\rm p}
&=
\eta_{y'}^{2} \langle v_{x'}^{2}\rangle_{\rm p}
+
\eta_{x'}^{2} \langle v_{y'}^{2}\rangle_{\rm p}
-
2\eta_{x'}\eta_{y'} \langle v_{x'}v_{y'}\rangle_{\rm p},
\label{eq:vpost2_raw}
\\
\langle v_{\rm POSr}v_{\rm POSt}\rangle_{\rm p}
&=
\eta_{x'}\eta_{y'} \left(\langle v_{x'}^{2}\rangle_{\rm p} - \langle v_{y'}^{2}\rangle_{\rm p}\right)
+
\left(\eta_{y'}^{2}-\eta_{x'}^{2}\right) \langle v_{x'}v_{y'}\rangle_{\rm p}.
\label{eq:vposr_vpost_raw}
\end{align}
\label{eq:posrt_raw_second_moments}
\end{subequations}
The corresponding projected dispersions are obtained by subtracting the
products of the projected first moments, as represented in equation~\ref{eq:projected_dispersion_from_raw}.

Finally, if additive velocity zero points are fitted in the projected
Cartesian components, their contribution to the projected polar
components is also position-dependent:
\begin{subequations}
\begin{align}
v_{0,{\rm POSr}}
&=
\eta_{x'} v_{0,x'}
+
\eta_{y'} v_{0,y'},
\\
v_{0,{\rm POSt}}
&=
\eta_{y'} v_{0,x'}
-
\eta_{x'} v_{0,y'}.
\end{align}
\label{eq:posrt_zero_points}
\end{subequations}
% Therefore, constant zero points in $v_{x'}$ and $v_{y'}$ do not
% generally correspond to constant zero points in
% $v_{\rm POSr}$ and $v_{\rm POSt}$.

\subsection{Likelihood implementation}
\label{ssec:rotation_likelihood_implementation}

The fits use the projected first and second moments above to construct,
for each star, a multivariate Gaussian model in the observed velocity
components. For a star \(j\), let \(\mathbf{v}_j\) be the selected
observed velocity vector, drawn from any chosen subset of
\(\{v_{z'},v_{y'},v_{x'}\}\). The model predicts the corresponding mean
vector \(\boldsymbol{\mu}_j\) from
equations~\ref{eq:vzprime_final_compact}--\ref{eq:vxprime_final_compact}
and the covariance matrix \(\mathbf{C}_j\) from
equations~\ref{eq:vxprime2_projected_TNM}--\ref{eq:vyprime_vzprime_projected_TNM}
after applying equations~\ref{eq:projected_dispersion_from_raw} and
\ref{eq:projected_covariance_from_raw}. The likelihood is then
\begin{equation}
    -\ln\mathcal{L}
    =
    \frac{1}{2}
    \sum_{j}
    \left[
    k\ln(2\pi)
    +
    \ln\det\mathbf{C}_j
    +
    \mathbf{r}_j^{\rm T}
    \mathbf{C}_j^{-1}
    \mathbf{r}_j
    \right],
    \label{eq:rotation_multivariate_likelihood}
\end{equation}
where \(k\) is the number of fitted velocity components and
\begin{equation}
    \mathbf{r}_j
    =
    \boldsymbol{\mu}_j+\mathbf{v}_0-\mathbf{v}_j.
\end{equation}
Here, \(\mathbf{v}_0\) is a vector of additive velocity zero-point
shifts, fitted independently for each included velocity dimension. These
zero points shift the model means only; they do not modify the
covariance matrices. A small diagonal jitter is added to \(\mathbf{C}_j\)
only as a numerical safeguard against nearly singular matrices.

The structural parameters of the projected Plummer profile are taken
from the fits of \citet{Vitral+26} to the same simulated data: the centre, scale radius
\(r_{\rm scale}\), projected axial ratio \(q_{\rm p}\), and photometric
position angle. The kinematic fit then varies the rotation amplitude
\(\omega\), the inverse-temperature parameter \(\mu\), and the
transition radius \(r_{\phi}\), using flat priors in
\(\log_{10}\omega\), \(\log_{10}\mu\), and \(\log_{10}r_{\phi}\).
The inclination is optimised through \(u=\cos i\), with
\(0\le u\le q_{\rm p}\), which is equivalent to enforcing a real
intrinsic flattening through
\begin{equation}
    q^{2}
    =
    \frac{q_{\rm p}^{2}-\cos^{2}i}{\sin^{2}i}.
\end{equation}
The projected kinematic axis angle $\xi_{\rm rot}$, which defines the $(x',y')$ frame used in the equations above, is also fitted. The sign of the projected rotation is therefore captured by the fitted axis angle, while $\omega$ is kept positive. Allowing $\xi_{\rm rot}$ to differ from the photometric position angle means that the surface and volume densities entering the projection integrals of equation~\ref{eq:projected_mean_definition} are effectively allowed to be misaligned with the photometric ellipsoid, even though the model retains the corresponding scale radius and projected axial ratio.
This introduces a controlled internal inconsistency whenever $\xi_{\rm rot}$ differs from the photometric position angle. This inconsistency notwithstanding, it is precisely the kind of departure from a smooth axisymmetric kinematic configuration that we aim to probe with the current formalism.

In practice, the integrals \(\mathcal{T}_{N}^{(M)}\) are evaluated with
a vectorised Gauss--Legendre quadrature rule over \(t\in[-1,+1]\). The optimisation is performed with Markov Chain Monte Carlo sampling using the \textsc{emcee} package \citep{ForemanMackey+13}. We run each fit for 20,000 steps with $2 \, N_{\rm param}+1$ walkers, where $N_{\rm param}$ is the number of fitted parameters, and discard the first half of each chain as burn-in.
The resulting posterior samples are then used to estimate parameter uncertainties. For non-angular parameters, we summarise the ensemble of successful fits using the median and uncertainties based on 16th--84th percentiles. For angular parameters such as $i$ and $\xi_{\rm rot}$, we instead use directional statistics \citep{Mardia&Jupp99}: the central value is computed from the mean direction of the unit vectors associated with each fitted angle, and the uncertainty is estimated from the circular median absolute deviation \citep[cf. equation 2 from][]{Vitral+26-sculptor} around that direction. This avoids artificial discontinuities when fitted angles lie close to the boundaries of their periodic domains.

\subsection{Forward mock realisations}
\label{ssec:rotation_mock_realisations}

For visualisation and diagnostic purposes, we also construct forward
Monte-Carlo realisations of the fitted six-dimensional model. First, we
sample positions from the intrinsic axisymmetric Plummer density by
drawing a spherical Plummer distribution in the auxiliary coordinate
\((x,y,z/q)\) and then compressing the third coordinate by the intrinsic
axis ratio \(q\). This produces the density profile of
equation~\ref{eq:intrinsic_plummer_density}.

At each sampled position, we draw velocities from the local
Lynden-Bell Gaussian distribution function. In spherical coordinates, this means
sampling
\begin{equation}
    v_r\sim\mathcal{N}(0,\sigma_r^{2}),
    \qquad
    v_{\theta}\sim\mathcal{N}(0,\sigma_{\theta}^{2}),
    \qquad
    v_{\phi}\sim
    \mathcal{N}(\langle v_{\phi}\rangle,\sigma_{\phi}^{2}),
\end{equation}
with \(\langle v_{\phi}\rangle\) given by
equation~\ref{eq:rotation_law} and the dispersions implied by
\citet[][equation~56]{LyndenBell67}. The sampled velocities are then
rotated into intrinsic Cartesian coordinates, projected to the
\((x',y',z')\) frame using the fitted inclination, and finally rotated
to the observed sky frame using the fitted \(\xi_{\rm rot}\). Finally, the
fitted velocity zero points are added to the mock projected velocities.

Importantly, these mocks are not used to define the likelihood. Instead, they are used only as
forward diagnostic realisations of the fitted model, allowing the
observed local velocity maps to be compared with a noisy projected
realisation of the same intrinsic Lynden-Bell--Plummer distribution.

\subsection{Recovery tests with mock data}
\label{ssec:rotation_recovery_tests}

We tested the numerical implementation of the likelihood described above by applying the full fitting routine to mock data sets generated from known input parameters. In each case, the mock catalogue was constructed from the same projected Lynden-Bell--Plummer model used in the fit, and the recovered posterior summaries were compared directly to the true values. For stellar samples with $N_{\star}\sim\mathcal{O}\left(10^{5}\right)$, all fitted parameters were recovered within the quoted $3\sigma$ uncertainties, with relative errors typically below $5$ per cent. This indicates that, in the high-count regime, the adopted quadrature, projection, and MCMC machinery do not introduce measurable biases in the recovered rotation parameters.

We repeated the same exercise for lower tracer counts in order to assess the sampling regime in which the model parameters remain identifiable. For $N_{\star}\sim\mathcal{O}\left(10^{4}\right)$, the recovery remained broadly unchanged for most parameters, with the exception of the inclination, which showed relative deviations from the expected value closer to $10$ per cent, preferring higher, more edge-on configurations. For $N_{\star}\sim\mathcal{O}\left(10^{3}\right)$, additional parameters could become poorly constrained depending on the input parameter set, including the rotation-amplitude parameter $\omega$ and the position angle of the rotation axis, $\xi_{\rm rot}$, although not typically all at once. These tests therefore suggest that the fitting procedure is reliable for sufficiently large stellar samples, but that deprojection-sensitive quantities should be interpreted with caution as the number of tracers decreases.

Encouragingly, this degradation does not affect all derived quantities equally: in particular, the dimensionless rotation scale $(v/\sigma)$, which combines $\omega$, $r_{\phi}$, and $\mu$ through equation~\eqref{eq:lb_v_over_sigma}, remains well constrained within fractions of $\sigma$ even for samples with $N_{\star}\sim\mathcal{O}\left(10^{3}\right)$. This is encouraging for observational applications, as many dwarf-galaxy data sets contain at most a few thousand stars with well-measured line-of-sight kinematics. While individual model parameters may become partially degenerate in this regime, their combination into a $(v/\sigma)$-like observable can still provide a robust summary of the relative importance of ordered and random motion.

\section{Handling of spectral features}
\label{sec:spectral_features}

This appendix summarises the numerical procedure used to construct and fit the
velocity-fluctuation spectra analysed in the main text. Throughout this appendix we consider
one projected velocity component at a time, $u\in\{x',y', {\rm POSr}, {\rm POSt}, z'\}$, where $z'$ is the
line-of-sight direction (cf. equation~\ref{eq:zprime-los-definition}).

\subsection{NUFFT power-spectra and azimuthal averaging}
\label{ssec:nufft_power_spectra}

The stars in our models do not, in general, populate every cell of a regular, well-sampled Cartesian grid, especially at large projected radii. We therefore estimate the Fourier amplitudes of the discrete $\chi_u$ field using a \textit{non-uniform} fast Fourier transform \citep*{Barnett+19,Barnett20}. Before doing so, the projected coordinates are centred, rotated into the adopted analysis frame, and expressed in units of the fitted Plummer scale radius $r_{\rm scale}$, following the same convention as in Paper~I.

We retain only stars within the square window
\begin{equation}
-L \leq X_j < L,
\qquad
-L \leq Y_j < L,
\qquad
L=5.2,
\label{eq:nufft_box_definition}
\end{equation}
where $X$ and $Y$ are dimensionless coordinates expressed in units of $r_{\rm scale}$. This choice avoids using the most extended regions of the stellar distribution, which in real data-like applications are more likely to be affected by external perturbations, tidal debris, or contamination from field stars.

The mean of the retained $\chi_u$ values is subtracted before transforming, removing the zero-frequency mode so that the measured spectrum describes fluctuations about the mean residual field. The Fourier
amplitude at wave vector $\mathbf{k}=(k_x,k_y)$ is therefore
\begin{equation}
    \widehat{\chi}_{u}(\mathbf{k})
    =
    \frac{1}{N_{\rm in}}
    \sum_{j=1}^{N_{\rm in}}
    \left(\chi_{u,j}-\overline{\chi}_{u}\right)
    \exp\left[-2\pi i\left(k_x X_j+k_y Y_j\right)\right],
    \label{eq:nufft_chi_amplitude}
\end{equation}
where $N_{\rm in}$ is the number of retained stars. Frequencies are reported in
cycles per $r_{\rm scale}$ rather than angular-frequency units. In practice, the
coordinate mapping supplied to the NUFFT is $\theta_X=\pi X/L$ and
$\theta_Y=\pi Y/L$, while the Fourier-mode indices $m,n$ correspond to
$k_x=m/(2L)$ and $k_y=n/(2L)$.

The number of Fourier modes is chosen adaptively from the local sampling of the
stellar distribution. We compute a reference nearest-neighbour spacing
$d_{\rm ref}$ from the retained points, using the median distance to the
$k$-th neighbour, with $k=\max(1,\lfloor N_{\rm in}/100\rfloor)$ unless specified
otherwise. The trusted wavelength is then
$\lambda_{\rm trust}=d_{\rm ref}$, and the number of modes is chosen as the nearest allowed
even value to $4L/\lambda_{\rm trust}$, subject to the bounds used in the
analysis. This gives an effective minimum wavelength
\begin{equation}
    \lambda_{\rm min}=\frac{4L}{N_{\rm modes}},
    \qquad
    k_{\rm max}=\lambda_{\rm min}^{-1}.
    \label{eq:nufft_resolution_definition}
\end{equation}
The two-dimensional power is then defined as
\begin{equation}
    P_{u}(k_x,k_y)
    =
    \left|\widehat{\chi}_{u}(k_x,k_y)\right|^{2}.
    \label{eq:nufft_power_2d}
\end{equation}

For the one-dimensional summaries used in the main text, and following Paper~I, we azimuthally average the two-dimensional power-spectrum as a function of
\begin{equation}
\kappa = \left(k_x^2+k_y^2\right)^{1/2}.
\end{equation}
This reduces the spectrum to a radial profile in Fourier space, making the characteristic scales of any excess power easier to interpret, although admittedly diluting localised azimuthal features.
The $\kappa$ values are divided into $N_{\kappa}=60$ linearly spaced bins between
zero and the largest sampled radial frequency. The saved spectrum is the pair
\begin{equation}
    \left\{\kappa_b,\mathcal{S}_{\chi_u\chi_u}(\kappa_b)\right\},
    \label{eq:saved_power_spectrum_pair}
\end{equation}
where $\kappa_b$ is the centre of bin $b$, and
$\mathcal{S}_{\chi_u\chi_u}(\kappa_b)$ is the mean of $P_u(k_x,k_y)$ over all
Fourier modes whose radial frequency falls in that bin. Empty, non-finite, zero,
or negative bins are naturally discarded.

\subsection{Model selection for spectral features}
\label{ssec:spectral_model_selection}

We fit each azimuthally averaged spectrum with two candidate models. The null model contains only a constant high-frequency floor, which captures statistical fluctuations around a smooth baseline,
\begin{equation}
M_0(\kappa)=\mathcal{C},
\label{eq:spectral_model_m0}
\end{equation}
whereas the one-feature model adds a single Voigt component,\footnote{Unlike in Paper~I, we find that a single Voigt component provides a satisfactory description of the residual power from subhalo-induced perturbations.}
\begin{equation}
M_1(\kappa)
=
\mathcal{C}
+
f\,V\left(\kappa, \, \mu_{\kappa}, \, \sigma, \, \gamma\right).
\label{eq:spectral_model_m1}
\end{equation}

Here, $\mathcal{C}$ is the noise floor, $f$ is the integrated Voigt flux,
$\mu_{\kappa}$ is the central frequency of the feature, and
$V(\kappa, \, \mu_{\kappa}, \, \sigma, \, \gamma)$ is the Voigt profile. In practice, we implement the latter parameters as
\begin{equation}
    \sigma=q \, \delta,
    \qquad
    \gamma=(1-q) \, \delta,
    \qquad
    0\leq q\leq 1.
    \label{eq:voigt_width_decomposition}
\end{equation}
The fitted parameters are therefore $\log_{10}\mathcal{C}$ for $M_0$, and
$\{\log_{10}\mathcal{C}, \, \log_{10}f, \, \log_{10}\mu_{\kappa}, \, \log_{10}\delta, \, q\}$ for
$M_1$. The fits are performed in logarithmic power, by minimising the residuals
between $\log_{10}\mathcal{S}_{\chi_u\chi_u}$ and $\log_{10}M_i$. For the
$M_1$ fit, the constant floor is constrained to remain within the uncertainty
range inferred from the $M_0$ floor fit, preventing the Voigt component from
absorbing noisy features from the low-frequency noise level.

The preferred model is selected using the Bayesian information criterion (BIC, \citealt{Schwarz78}),\footnote{We prefer this metric to Akaike-based criteria \citep[e.g., ][]{Akaike73,Sugiyara78} because it penalises the $M_{1}$ model more strongly against spurious spectral noise that might mimic a peak in the power-spectrum, especially at lower spatial frequencies that are more poorly sampled.}
computed from the log-power residual sum of squares,
\begin{equation}
    {\rm BIC}_i
    =
    N_{\rm data}\ln\left(\frac{{\rm RSS}_i}{N_{\rm data}}\right)
    +
    N_{\mathrm{param}, \, i} \ln N_{\rm data},
    \label{eq:spectral_bic_definition}
\end{equation}
where $N_{\rm data}$ is the number of fitted $\kappa$ bins, $N_{\mathrm{param}, \, i}$ is the number of free
parameters in model $M_i$, and ${\rm RSS}_i$ is evaluated in $\log_{10}$ power.
This comparison penalises narrow noise fluctuations that would otherwise be
absorbed by a Voigt profile. We also impose two explicit guards on the $M_1$
component. First, the effective Voigt width \citep{Whiting68,Olivero&Longbothum77},
\begin{equation}
    w \equiv \frac{\delta}{2} \, \left[ 1 - q + \sqrt{1 + q \, (5 \, q - 2)} \right],
    \label{eq:effective_voigt_width}
\end{equation}
must be at least one local resolution element, estimated from the neighbouring
$\kappa$-bin spacing at $\mu_{\kappa}$. Second, the component must be contained
within the fitted spectral range, such that
$\mu_{\kappa}+w< {\rm max}(\{\kappa_j\})$. Fits violating either condition are
assigned a non-finite information criterion and are therefore rejected. These
criteria suppress unresolved bin-to-bin fluctuations and edge-truncated peaks,
which are especially easy to misidentify in noisy spectra.

\section{Projection of triaxial shapes and kinematic misalignments}
\label{sec:triaxial_projection}

Here, we describe the geometrical prescription used to predict the joint
distribution of apparent ellipticity and photometric--kinematic misalignment
for a triaxial stellar system. We follow the definitions of
\citet{Franx+91}, but evaluate the projection directly in Cartesian
coordinates rather than through the conical-coordinate formalism of their
Appendix~A.

We consider a density distribution stratified on similar, co-aligned
ellipsoids,
\begin{equation}
    m^{2}
    =
    \frac{x^{2}}{a^{2}}
    +
    \frac{y^{2}}{b^{2}}
    +
    \frac{z^{2}}{c^{2}},
    \qquad
    a\geq b\geq c,
    \label{eq:triaxial_ellipsoid}
\end{equation}
where the $x$-, $y$-, and $z$-axes are the intrinsic long, intermediate,
and short axes, respectively. 
Following \citet{Franx+91}, we characterise the intrinsic shape through
\begin{equation}
    \epsilon_{1}=1-\frac{c}{a},
    \qquad
    \epsilon_{2}=1-\frac{b}{a},
    \label{eq:triaxial_intrinsic_eps}
\end{equation}
and the triaxiality parameter as
\begin{equation}
    T
    =
    \frac{a^{2}-b^{2}}{a^{2}-c^{2}}
    =
    \frac{\epsilon_{2}(2-\epsilon_{2})}
         {\epsilon_{1}(2-\epsilon_{1})}.
    \label{eq:triaxial_T}
\end{equation}
Since the overall scale is irrelevant for projection, we set $a=1$, such
that $b=1-\epsilon_{2}$ and $c=1-\epsilon_{1}$.

% The overall scale is irrelevant for projection, and we therefore set $a=1$, such that $b=1-\epsilon_{2}$ and $c=1-\epsilon_{1}$.
% Although the apparent position-angle distribution depends only on $T$, the joint distribution in apparent ellipticity and position angle also depends on the intrinsic flattening; we therefore retain $(\epsilon_{1},\epsilon_{2})$ explicitly.

For a viewing direction specified by the usual polar angles
$(\theta,\phi)$, the line-of-sight unit vector is
\begin{equation}
    \mathbf{n}
    =
    \left(
    \sin\theta\cos\phi,\,
    \sin\theta\sin\phi,\,
    \cos\theta
    \right).
    \label{eq:triaxial_los}
\end{equation}
We define an orthonormal basis in the plane of the sky as
\begin{align}
    \mathbf{u}
    &=
    \left(
    -\sin\phi,\,
    \cos\phi,\,
    0
    \right),
    \nonumber\\
    \mathbf{v}
    &=
    \mathbf{n}\times\mathbf{u}
    =
    \left(
    -\cos\theta\cos\phi,\,
    -\cos\theta\sin\phi,\,
    \sin\theta
    \right).
    \label{eq:triaxial_sky_basis}
\end{align}
This particular basis is convenient because $\mathbf{v}$ is parallel to
the projection of the intrinsic short axis onto the sky. Indeed,
\begin{equation}
    \hat{\mathbf{z}}_{\perp}
    =
    \hat{\mathbf{z}}
    -
    (\hat{\mathbf{z}}\cdot\mathbf{n})\mathbf{n}
    =
    \sin\theta\,\mathbf{v}.
    \label{eq:projected_short_axis}
\end{equation}

To obtain the apparent shape, we write
$\mathbf{D}=\mathrm{diag}(a,b,c)$. The ellipsoid in
equation~\eqref{eq:triaxial_ellipsoid} can then be regarded as the image
of the unit sphere under the transformation
$\mathbf{r}=\mathbf{D}\, \mathbf{s}$, with $|\mathbf{s}|\leq1$.
Projecting $\mathbf{r}$ onto the $(\mathbf{u},\mathbf{v})$ sky plane
therefore maps the unit sphere onto an ellipse. 
From this, one can show, through singular value decomposition, that the squared semi-axes of the apparent ellipse, viewed after projection, are the eigenvalues of the corresponding $2\times2$ projected
matrix
\begin{equation}
    \mathbf{Q}
    =
    \begin{pmatrix}
        \mathbf{u}^{\rm T}\mathbf{D}^{2}\mathbf{u}
        &
        \mathbf{u}^{\rm T}\mathbf{D}^{2}\mathbf{v}
        \\
        \mathbf{v}^{\rm T}\mathbf{D}^{2}\mathbf{u}
        &
        \mathbf{v}^{\rm T}\mathbf{D}^{2}\mathbf{v}
    \end{pmatrix},
    \label{eq:triaxial_projected_matrix}
\end{equation}
whose elements are explicitly
\begin{align}
    Q_{11}
    &=
    a^{2}\sin^{2}\phi+b^{2}\cos^{2}\phi,
    \nonumber\\
    Q_{22}
    &=
    \cos^{2}\theta
    \left(
    a^{2}\cos^{2}\phi+b^{2}\sin^{2}\phi
    \right)
    +
    c^{2}\sin^{2}\theta,
    \nonumber\\
    Q_{12}
    &=
    (a^{2}-b^{2})
    \cos\theta\sin\phi\cos\phi.
    \label{eq:triaxial_projected_matrix_components}
\end{align}
The two eigenvalues are
\begin{equation}
    \lambda_{\pm}
    =
    \frac{1}{2}
    \left[
    Q_{11}+Q_{22}
    \pm
    \sqrt{
    (Q_{11}-Q_{22})^{2}+4Q_{12}^{2}
    }
    \right],
    \label{eq:triaxial_eigenvalues}
\end{equation}
where $\sqrt{\lambda_{+}}$ and $\sqrt{\lambda_{-}}$ are the apparent
major and minor semi-axes, respectively. The apparent ellipticity is
therefore
\begin{equation}
    \epsilon
    =
    1-
    \sqrt{\frac{\lambda_{-}}{\lambda_{+}}}.
    \label{eq:triaxial_apparent_eps}
\end{equation}

The orientation of the apparent ellipse follows from the eigenvectors of
$\mathbf{Q}$. Writing the apparent major-axis direction in the sky basis as
\begin{equation}
    \mathbf{e}_{\rm maj}
    =
    \cos\alpha\,\mathbf{u}
    +
    \sin\alpha\,\mathbf{v},
\end{equation}
the rotation angle $\alpha$ that diagonalises $\mathbf{Q}$ is
\begin{equation}
    \alpha
    =
    \frac{1}{2}
    \operatorname{atan2}
    \left(
    2Q_{12},\,Q_{11}-Q_{22}
    \right),
    \label{eq:triaxial_alpha}
\end{equation}
where the first argument of the $\operatorname{atan2}$ function represents the opposite side of the right triangle.
Thus, $\alpha$ is the angle between the apparent major axis and $\mathbf{u}$. Since the apparent minor axis is perpendicular to the apparent major axis, while $\mathbf{v}$ is perpendicular to $\mathbf{u}$, the angle between the apparent minor axis and $\mathbf{v}$ is the same $\alpha$.
Because $\mathbf{v}$ is parallel to the projected intrinsic short axis
(equation~\ref{eq:projected_short_axis}), this angle is precisely the
photometric position-angle offset, $\Gamma_{\rm minor}=|\alpha|$,
up to the usual folding of position angles into the interval
$0\leq\Gamma_{\rm minor}\leq\pi/2$.

As in the perfectly aligned models of \citet{Franx+91} considered here,
we assume that the intrinsic angular momentum is parallel to the
intrinsic short axis. Its projection is therefore parallel to
$\mathbf{v}$, giving $\Gamma_{\rm kin}=0$. The apparent kinematic
misalignment that we here denote $|\Delta \xi|$ consequently reduces to
\begin{equation}
    |\Delta \xi| \equiv
    \Psi
    =
    |\Gamma_{\rm minor}|
    =
    |\alpha|.
    \label{eq:triaxial_apparent_psi}
\end{equation}
Thus, even a perfectly aligned intrinsic angular momentum can show a
non-zero apparent photometric--kinematic misalignment as a consequence
of projecting a triaxial figure. 
% More general models with an intrinsically
% misaligned angular momentum require an additional intrinsic
% misalignment angle and cannot be specified by
% $(\epsilon_{1},\epsilon_{2})$ alone.

We finally construct the joint probability distribution
$p(\epsilon,\Psi)$ by averaging over isotropically distributed viewing
directions, i.e. rather than evaluating the analytic Jacobian in Appendix~A
of \citet{Franx+91}, we perform the equivalent forward projection over
the unit sphere.
% For the density calculation, we use $N=5\times10^{6}$
% quasi-uniform directions generated with a Fibonacci sphere,
% \begin{equation}
%     \cos\theta_i
%     =
%     1-\frac{2(i+1/2)}{N},
%     \qquad
%     \phi_i
%     =
%     i\pi(3-\sqrt{5})
%     \pmod{2\pi},
%     \label{eq:fibonacci_sphere}
% \end{equation}
% which provides an approximately equal-area sampling of the isotropic
% measure $d\Omega=\sin\theta\,d\theta\,d\phi$. Each direction is mapped
% through equations~\eqref{eq:triaxial_apparent_eps} and
% \eqref{eq:triaxial_apparent_psi} into one $(\epsilon,\Psi)$ pair.
% For bins of widths $\Delta\epsilon$ and $\Delta\Psi$, the probability
% density is estimated as
% \begin{equation}
%     p(\epsilon_j,\Psi_k)
%     \simeq
%     \frac{N_{jk}}
%          {N\,\Delta\epsilon\,\Delta\Psi},
%     \label{eq:triaxial_probability_density}
% \end{equation}
% where $N_{jk}$ is the number of viewing directions falling in the
% corresponding bin. In practice, we use 100 bins over
% $0\leq\epsilon\leq\epsilon_1$ and 100 bins over
% $0\leq\Psi\leq\pi/2$. The resulting density is mildly smoothed with a
% Gaussian kernel of width $1.5$ bins and renormalised to unit integral.
% The individual points shown in Figure~\ref{fig:triaxiality-misalignment}
% are generated independently from random isotropic orientations, using
% $\cos\theta\sim\mathcal{U}(-1,1)$ and
% $\phi\sim\mathcal{U}(0,2\pi)$
In practice, for each simulation snapshot, we first propagate the uncertainty in its
intrinsic shape by drawing $N_{1} = 10$ realisations of
$(\epsilon_{1},\epsilon_{2})$ from the corresponding triaxial fit. For each of these intrinsic-shape realisations, we then draw $N_{2} = 100$ independent isotropic viewing directions, using
$\cos\theta\sim\mathcal{U}(-1,1)$ and
$\phi\sim\mathcal{U}(0,2\pi)$, and project the system according to
equations~\eqref{eq:triaxial_apparent_eps} and
\eqref{eq:triaxial_apparent_psi}. Each snapshot therefore contributes
$N_{1} N_{2}$ samples in the $(\epsilon,\Psi)$ plane.

Rather than adopting a single intrinsic shape averaged over the full
simulation, we repeat this procedure independently for every snapshot
and superpose the resulting projected samples.
The final theoretical distribution is therefore an empirical mixture of
the aligned predictions associated with the full set of intrinsic
shapes explored by the simulation, while also propagating the
uncertainties of the individual triaxial fits. The black points in
Figure~\ref{fig:triaxiality-misalignment} are drawn from this combined
sample, and the corresponding density contours are estimated directly
from the same pooled $(\epsilon,\Psi)$ distribution.

\section{Three-dimensional angular-momentum coherence}
\label{sec:angular_momentum_coherence}

Here, we describe the three-dimensional angular-momentum diagnostics used
to assess whether the simulated stellar systems develop a preferred
streaming direction, independently of the projected Lynden-Bell--Plummer
rotation model introduced in Appendix~\ref{sec:rotation_models}.

\subsection{Angular-momentum coherence}

For each snapshot, we adopt the centre $\mathbf{x}_{0}$ inferred from the
corresponding three-dimensional triaxial Plummer fit and define the
position of star $i$ relative to this centre as
\begin{equation}
    \mathbf{r}_{i}
    =
    \mathbf{x}_{i}-\mathbf{x}_{0}.
\end{equation}
For a trial bulk velocity $\mathbf{v}_{\rm bulk}$, the specific angular momentum of star $i$, and its direction are then
\begin{equation}
    \mathbf{j}_{i}
    =
    \mathbf{r}_{i}
    \times
    \left(
    \mathbf{v}_{i}-\mathbf{v}_{\rm bulk}
    \right),
    \qquad
    \hat{\mathbf{j}}_{i}
    =
    \frac{\mathbf{j}_{i}}{|\mathbf{j}_{i}|}.
    \label{eq:specific_angular_momentum}
\end{equation}

We characterise the degree of common streaming around a trial unit axis
$\hat{\mathbf{n}}$ through two complementary angular-momentum coherence quantities:
\begin{equation}
    C
    =
    \left|
    \frac{
    \sum_{i}
    \hat{\mathbf{n}}\cdot\mathbf{j}_{i}
    }{
    \sum_{i}|\mathbf{j}_{i}|
    }
    \right|, \qquad
    R
    =
    \left|
    \frac{1}{N}
    \sum_{i}
    \hat{\mathbf{n}}\cdot\hat{\mathbf{j}}_{i}
    \right|.
    \label{eq:Cj_definition}
\end{equation}
While $C$ gives greater weight to stars carrying larger specific angular
momentum, $R$ considers only the directions of the individual
angular-momenta, thus being more robust against outliers having high $|\mathbf{j}_{i}|$ values.
Both quantities lie between zero and unity: values approaching
unity indicate angular momenta concentrated around a common direction,
whereas low values correspond to weak or mutually cancelling streaming.

The preferred axis and bulk velocity are determined jointly from these
two diagnostics. Specifically, we maximise
\begin{equation}
    \mathcal{S}
    =
    C\,R,
    \label{eq:angular_momentum_coherence_score}
\end{equation}
over the two angular degrees of freedom defining
$\hat{\mathbf{n}}$ and the three Cartesian components of
$\mathbf{v}_{\rm bulk}$. Since both $C$ and $R$ are axial
magnitudes, the transformation
$\hat{\mathbf{n}}\rightarrow-\hat{\mathbf{n}}$
otherwise generates an equivalent solution. We remove this degeneracy
by requiring $\hat{\mathbf{n}}\cdot \sum_i\mathbf{j}_i > 0$.
In practice, we first locate the maximum of
equation~\eqref{eq:angular_momentum_coherence_score} numerically and
then explore its neighbourhood with MCMC sampling. 
% Since
% $\mathcal{S}$ is a descriptive coherence statistic rather than
% a generative likelihood, the MCMC is based on the corresponding
% generalised score posterior,
% \begin{equation}
%     \ln \mathcal{P}_{\rm score}
%     =
%     {\rm const}
%     +
%     N_{\rm eff}\,\mathcal{S},
%     \label{eq:angular_momentum_score_posterior}
% \end{equation}
% within the adopted parameter bounds and subject to
% equation~\eqref{eq:angular_momentum_axis_sign}. Here,
% $N_{\rm eff}$ is the number of stars with a well-defined angular-momentum
% direction. The resulting sampling is therefore used to identify and
% characterise the preferred streaming axis and bulk velocity, rather
% than being interpreted as a physical likelihood for the stellar
% phase-space distribution.

% With the fitted orientation convention, both coherence statistics are
% non-negative and satisfy
% \begin{equation}
%     0\leq C\leq1,
%     \qquad
%     0\leq R\leq1.
% \end{equation}
% Values approaching unity indicate that the stellar angular momenta are
% strongly concentrated around a common direction, while values close to
% zero indicate little net preference for streaming about the fitted
% axis.

\subsection{Sampling uncertainties on the coherence statistics}
\label{ssec:angular_momentum_coherence_uncertainties}

The dispersion of the individual stellar angular-momentum directions
is generally much broader than the uncertainty on the snapshot-level
coherence measurement.\footnote{Analogously, the uncertainty around the mean of a distribution is generally smaller than the standard deviation of that same distribution.} We therefore distinguish the intrinsic
star-to-star distribution of angular momenta from the sampling uncertainty on
$C$ and $R$, and use the latter when comparing the coherence
statistics between snapshots.
The sampling uncertainty of $C$ is therefore estimated as
\begin{equation}
    \sigma_{C}
    =
    \frac{1}
         {\sqrt{N}\,\overline{|\mathbf{j}|}}
    \left[
    \frac{1}{N-1}
    \sum_i
    \left(
        \hat{\mathbf{n}}\cdot\mathbf{j}_i
        -
        C|\mathbf{j}_i|
    \right)^2
    \right]^{1/2}.
    \label{eq:Cj_sampling_uncertainty}
\end{equation}
This expression accounts for both the variation in the projected
angular momentum and the unequal angular-momentum amplitudes carried
by the individual stars.

For the directional statistic, the corresponding uncertainty is simply the
standard error of their mean,
\begin{equation}
    \sigma_R
    =
    \left[
    \frac{1}{N(N-1)}
    \sum_i
    \left(\hat{\mathbf{n}}\cdot\hat{\mathbf{j}}_i-R\right)^2
    \right]^{1/2}.
    \label{eq:R_sampling_uncertainty}
\end{equation}
For this expression, $N$ denotes the number of stars for which the
direction of $\mathbf{j}_i$ is numerically well defined. 
% An overall reversal of the angular-momentum axis does not affect either of these uncertainties.

These $\sigma_{C}$ and $\sigma_R$ values quantify the finite-sampling
uncertainty of the global coherence diagnostics. They should therefore
not be confused with the 16th--84th percentile ranges of the individual
stellar distributions, which instead quantify the
intrinsic spread of angular-momentum directions within a snapshot, and can be larger as the system becomes more pressure-supported.

\subsection{Correlation fits}
\label{ssec:angular_momentum_coherence_correlations}

To quantify trends between the angular-momentum coherence and the
rotation diagnostics inferred from the Lynden-Bell--Plummer fits, we
use orthogonal distance regression (ODR) as implemented in the corresponding \textsc{scipy/Python} package. Unlike an ordinary
least-squares fit, ODR allows the uncertainties of both coordinates to
contribute to the inferred relation. For each pair of quantities we
fit $Y = mX+b$, where $(X,Y)$ denote the coordinates in which the corresponding panel
is displayed. Thus, for a logarithmic axis we use
$X=\log_{10}x$ or $Y=\log_{10}y$, together with the respective
transformed uncertainties, whereas linear axes are fitted directly in
the original quantities.
The ODR procedure then determines
the slope $m$ and intercept $b$ while allowing displacements along both
coordinates, and we use the corresponding ODR parameter uncertainty as
the quoted uncertainty on the slope. The slope parameter $m$, as defined above, is the one depicted in the legends of Figure~\ref{fig:correlations-streaming}.

Finally, we restrict these comparisons to snapshots for which the
relevant coherence and rotation signals are sufficiently well defined.
Whenever they enter a given correlation analysis, we exclude
measurements satisfying any of the following:
\begin{equation}
    C < 0.05,
    \qquad
    R < 0.05,
    \qquad
    (v/\sigma) < 0.05.
    \label{eq:coherence_correlation_thresholds}
\end{equation}
These low-amplitude measurements correspond to regimes in which the
preferred streaming direction or rotation amplitude becomes too weak
compared with the intrinsic stellar dispersion, and we therefore do
not interpret them as robust indicators of coherent rotation. The
threshold is applied to the snapshot measurements before
performing the ODR fits.

%%%%%%%%%%%%%%%%%%%%%%%%%%%%%%%%%%%%%%%%%%%%%%%%%%

% Don't change these lines
\bsp	% typesetting comment
\label{lastpage}
\end{document}